\documentclass[11pt,a4paper]{article}
\usepackage[english]{babel}

\usepackage[top=3cm,bottom=3cm,left=3cm,right=3cm,marginparwidth=0.75cm]{geometry}

\usepackage{multirow}
\usepackage{amsmath,amssymb,amsthm}
\usepackage{graphicx,xcolor}
\usepackage[colorlinks=true, allcolors=blue]{hyperref}
\usepackage{authblk}
\usepackage{natbib}

\DeclareMathOperator{\GP}{\mathrm{GP}}
\DeclareMathOperator{\Po}{\mathrm{Poi}}
\DeclareMathOperator{\Ga}{\mathrm{Gamma}}
\DeclareMathOperator{\Cov}{\mathrm{Cov}}
\DeclareMathOperator{\Var}{\mathrm{Var}}
\DeclareMathOperator{\NB}{\mathrm{NegBin}}

\newcommand{\cN}{\mathcal{N}}
\newcommand{\EE}{\mathbb{E}}

\usepackage{array}
\newcolumntype{L}[1]{>{\raggedright\arraybackslash}m{#1}}
\newcolumntype{C}[1]{>{\centering\arraybackslash}m{#1}}
\title{Analyzing Pension Fund Mortality  with Gaussian Processes in a Sub Population Framework}

\author[1]{Eduardo F. L. de Melo}
\author[2]{Michael Ludkovski}
\author[3]{Rodrigo S. Targino}
\affil[1]{School of Applied Mathematics, FGV, Rio de Janeiro, Brazil; SUSEP, Rio de Janeiro, Brazil; UERJ, Rio de Janeiro, Brazil}
\affil[2]{Department of Statistics and Applied Probability, University of California, Santa Barbara, California, USA}
\affil[3]{School of Applied Mathematics, FGV, Rio de Janeiro, Brazil}

\begin{document}

\maketitle

\ \ 

\begin{abstract}

\noindent Pension fund populations often have mortality experiences that are substantially different from the national benchmark. In a motivating case study of Brazilian corporate pension funds, pensioners are observed to have mortality that is 40-55\% below the national average, due to the underlying socioeconomic disparities. Direct analysis of a pension fund population is challenging due to very sparse data, with age-specific annual death counts often in low single digits. We design and study a collection of stochastic sub-population frameworks that coherently capture and project pensioner mortality rates via deflator factors relative to a reference population. Superseding parametric approaches, we propose Gaussian process (GP) based models that flexibly estimate Age- and/or Year-specific deflators. We demonstrate that the GP models achieve better goodness of fit and uncertainty quantification. Our models are illustrated on two Brazilian pension funds in the context of exogenous national and insurance industry mortality tables. The GP models are implemented in \textbf{R Stan} using a fully Bayesian approach and take into account over-dispersion relative to the Poisson likelihood.

\


\end{abstract}

\newpage

\section{Introduction}
The valuation of pension funds liabilities requires the use of consistent, realistic and up-to-date mortality rates and also the forecast of these mortality rates into the future for cash flow discount or asset-liability management. Such requirements are embedded in modern solvency assessment principles and financial reporting standards, see \cite{sandstrom2016handbook} and \cite{iasb2017ifrs17}. The vast majority of studies in mortality rates forecasting make use of national population data when performing applications, see \cite{Cairns2009, dowd2011gravity}; \cite{lee1992modeling, LI2005}; \cite{LI2013, renshaw2006cohort}, among others. 

Since national population is inevitably different from the pension fund population, there is an inherent basis risk in substituting the former mortality rates for the purposes of analyzing the latter. This basis risk may not present a meaningful problem when the difference between the national population and the underlying sub (selected) population is reasonable. For instance, in developed countries in Europe, the population mortality tends to be more homogeneous. However, in countries with high social inequalities, the income differences and healthcare access make the mortality rates significantly different among the selected and the overall national populations.

Brazil is a country with high social inequalities \citep{oecd2018brazil}. According to supervisory authorities, less than 8\% of Brazil's populace has a pension plan\footnote{\url{https://www.investidorinstitucional.com.br/sessoes/investidores/seguradoras/40265-previdencia-aberta-atinge-5-3-da-populacao-brasileira-em-maio.html}} and the fraction that does tends to be in the upper deciles of the national income distribution. As a result, national mortality rates are much higher than the ones for actuarial sub-populations, such as participants of pension funds or insurance customers.  Our data indicates that, on average, the mortality rates of the pension fund sub population are \emph{half} (49\%) of the mortality of the national Brazilian population for ages $60+$.

Mortality modeling for a pension fund must contend with two fundamental challenges. On the one hand, the population of interest is relatively small, so that stand-alone models lack sufficient credibility: there are simply not enough death counts to reliably estimate mortality rates. On the other hand, the pensioners often have materially distinct mortality patterns, so that simple adjustments of the national trends are inadequate. In this paper, we aim to systematically explore and ultimately bridge this gap. To this end, we construct a spectrum of \emph{deflator-based} stochastic models that relate the pension fund mortality to the reference population. As a starting point, we observe that the commonly used parametric deflators (such as deflators that are constant across ages) are too rigid and underfit. At the other extreme, allowing for independent age- (or year-) specific deflators leads to over-parametrization and overfitting. This motivates us to develop non-parametric deflators that are constrained to be smooth yet flexible enough. Specifically, we propose to employ Gaussian-process (GP)-based deflators, which naturally embed dependence while providing a fully-probabilistic, data-driven, non-parametric framework. 

Methodologically, we make three contributions. First, we provide a taxonomy of over half a dozen deflator models for small population mortality, systematically comparing them (including discussing multiple nesting relationships present) and offering a comprehensive ``menu'' for how to approach this problem. Second, we propose a new class of GP-driven deflators. Such models offer a valuable hybrid of leveraging the reference mortality table while building in sufficient flexibility. Third, with an emphasis on capturing model risk due to the small underlying population, we utilize throughout our analysis Bayesian methods, so that all our model hyperparameters come with priors. We employ the \texttt{Stan} package and underlying MCMC routines to provide uncertainty quantification, contributing to the literature on Bayesian mortality analysis. 

Empirically, we are motivated by and illustrate our developments with a dataset of two Brazilian pensioners. The pension fund in our primary case study (pension fund 1) consists of about 14,000 exposed lives with about 100 annual deaths and is compared to two national baselines. These pensioners have about half the mortality rate as those references, and notably exhibit a growing gap in longevity relative to the baselines. We demonstrate that our proposed models outperform extant alternatives, including the two extremes of our modeling spectrum: a stand-alone GP model, and a fully-specified fixed-deflator model. Given that age-specific death counts in this fund are often 0 or 1, we moreover explore the observation likelihood, namely the overdispersion in death counts relative to the Poisson baseline. Our analysis provides insights into the mortality experience in Brazil and in developing countries more broadly, offering new take-aways on pension funds in highly heterogeneous settings. To confirm the above findings, we further explore a second Brazilian pension fund that is slightly smaller.

Our interest in this paper is on ages 60 and over (60$+$), ages that play a relevant role for insurance and pension funds when paying annuities. This is another difference in our approach compared to the majority of papers in the mortality modeling literature. In this paper, Gaussian processes will be used in a subpopulation framework to forecast mortality at ages 60+ of small specific pension funds' populations, using as reference both annual Brazilian national mortality tables and a Brazilian insurance industry mortality table, which has a non-regular frequency of publication. The data used in this paper is thoroughly described in Section \ref{sec:data}.

The rest of the paper is organized as follows. Section \ref{sec:literature} reviews existing methods for sub-population models, as well as the Gaussian Process paradigm. Section \ref{sec:data} summarizes our Brazil pension fund dataset and the respective BR-EMS and IBGE reference mortality tables. Section \ref{sec:models} presents the list of the considered deflator models, both from existing approaches as well as new ones proposed here for the first time. Section \ref{sec:results} illustrates and discusses the respective fits for our motivating dataset, while Section \ref{sec:discussion} provides a statistical assessment across all the fitted models. Appendix summarizes the models fitted in this paper, additional plots for the primary pension fund and, finally, discussions for the second pension fund (pension fund 2).

\section{Sub population models}\label{sec:literature}

Given the small size of the modeled population, we directly work with death and exposed counts, disaggregated by age and calendar year. 
We employ the following notation
\begin{itemize} 
\item $d_{x,t}$ is the number of deaths at age $x$ and calendar year $t$ for the pension fund, where $x = \{60,...,89 \}$ and $t = \{2013,...,2019\}$; 
\item $E_{x,t}$ is the risk exposure at age $x$ and calendar year $t$ for the pension fund; 
\item $m^{ref}_{x,t}$ is the central mortality rate relative to age $x$ and calendar year $t$, where $ref =$ $\{BRA$ - Brazilian national mortality rates (IBGE), $IND$ - Brazilian insurance industry mortality rates (BR-EMS)$\}$.
\end{itemize} 

To overcome the scarce data available about the pension fund experience, we adopt the solution of external reference tables. The idea of such relational models is to adjust the reference table to the experience of the pension fund. In our work, we consider two  benchmark mortality tables for Brazil: $BRA$ also known as the IBGE (Brazilian Institute of Geography and Statistics) national population mortality table, and $IND$, aka BR-EMS, the insurance industry mortality table. These tables, as well as the pension fund's data, are described in detail in Section \ref{sec:data}. 

Reflecting the limited number of deaths in the pension fund population, the common  approach is to use a discrete (Poisson) likelihood where the reference mortality rate $m^{ref}_{x,t}$ is part of the intensity parameter:

\begin{align}\label{eq:deflator}
    d_{x,t} \sim \Po(e^{\theta_{x,t}}  m^{ref}_{x,t}  E_{x,t}),
\end{align}

\noindent and $e^{\theta_{x,t}}$ is the \emph{deflator} applied to the reference population. Model \eqref{eq:deflator} implies the fund mortality rate $m_{x,t} := \mathbb{E}[ d_{x,t}]/E_{x,t}$ is given by $m_{x,t}= e^{\theta_{x,t}} m^{ref}_{x,t}$. Hence, $$\theta_{x,t} = \log m_{x,t} - \log m^{ref}_{x,t}$$ is the difference on the log-scale between the fund mortality and the reference mortality, or equivalently the percent ratio between them. Since the funds' mortality is lower than the reference population's, $\theta_{x,t} < 0$.

Before describing our method in detail, we provide a review of existing deflator models. 
The deflator approach was introduced in \cite{hardy1998credibility}, who started with a constant deflator $d_{x,t} \sim \Po(\Theta \cdot m^{ref}_{x,t}  \cdot E_{x,t})$. Beyond directly fitting $\Theta$ as a static parameter, they consider a sequential Bayesian setup, where $\Theta$ is endowed with a prior, which is updated over time based on observed $(d_{x,t}, E_{x,t})$. Specifically, they leverage Poisson-Gamma conjugacy, to assign a Gamma prior  $\Theta \sim \Ga(c,c)$ which has mean 1 and  therefore \emph{a priori} the portfolio is expected to have the same mortality as the reference population. \cite{olivieri2012life} generalize to independent age-specific random effects, obtaining the posterior $\Theta_x | \mathcal{D} \sim \Ga( c + \sum_t d_{x,t}, c+ \sum_t m^{ref}_{x,t} E_{x,t})$.

\cite{olivieri2011stochastic} takes 
$$d_{x,t} \sim \Po(\Theta_{x,t} \cdot m^{ref}_{x,t}  \cdot E_{x,t}) \qquad\text{ where } \quad \Theta_{x,t} \sim \Ga( \alpha_{x,t}, \beta_{x,t})$$
and the hyperparameters $\alpha,\beta$ are sequentially updated over the years based on realized $d_{x,t}$.  \cite{vanberkum2017bayesian} use age-specific Bayesian ``random effects'', $d_{x,t} \sim \Po(\Theta_x \cdot  m^{ref}_{x,t}  \cdot  E_{x,t})$, where $m^{ref}$ is estimated using a Lee-Carter model, and the age-specific deflators $\Theta_x$ are given either independent $\Ga( c^i_x, c^i_x)$ priors (forcing $\mathbb{E}[\Theta_x ] =1$) or log-AR(1) structure, $\log \Theta_x = \mu + \rho \log \Theta_{x-1} + \eta$, with log-normal prior on $\log \Theta_{x_0}$, uniform prior on the variance of $\eta$ and logit-normal prior on $\rho$.

Making the deflators $\theta_x$ constant in time offers a coherent projection into the future. For instance, it is common to assume that the mortality trend of the small population (which is difficult to reliably estimate due to paucity of data) matches that of the reference, see e.g., \cite{olivieri2012life, salhi2015credibility, vanberkum2017bayesian, li2017coherent,hyndman2013coherent,salhi2017basis}. \cite{salhi2015credibility} focuses on time-dependent ``differential mortality law'', i.e.,$\theta_t$ depending only on year. 

Closely related to our work is \cite{tomas2015prospective}. They consider a sequence of models, starting with an endogenous nonparametric $d_{x,t} \sim \Po( E_{x,t} \cdot e^{f(x,t)})$ to the relational versions $d_{x,t} \sim \Po( E_{x,t} \cdot e^{f(\log m^{ref}_{x,t} )   } )$, $d_{x,t} \sim \Po(E_{x,t} \cdot m^{ref}_{x,t}  \cdot e^{f(x,t)} )$ where the smooth functions $f(x,t)$ are estimated using  local kernel weighted log-likelihood methods.

A related strand of literature directly proposes a parametric functional relationship between the mortality surfaces of the small and benchmark populations. In the original Brass model of \cite{brouhns2002poisson}, the authors take $m_{x,t}= \theta_1 (m^{ref}_{x,t})^{\theta_2}$ for parameters $\theta_1, \theta_2$ to be fitted by GLM. More generally, \cite{brouhns2002poisson} consider $m_{x,t} = \theta_0 f(m^{ref}_{x,t}) + \theta_1$, for a given link function $f$ (such as the logit). 
\cite{plat2009stochastic} directly models the ratio $P_{x,t} =(d_{x,t}/E_{x,t}) / (d^{ref}_{x,t}/E^{ref}_{x,t})$ of the observed portfolio and reference mortality rate, imposing a linear relationship, $P_{t,x} = a_t + b_t x + \epsilon_{x,t}$. \cite{Bienvenue2012} use distortion of probability distributions to connect $m^{ref}$ and $m$.

A different way to frame the small population problem is via a multi-population approach. Such joint modeling of the larger and smaller sub-population is especially appropriate when the smaller sub-population is sufficiently large and moreover, when starting with raw data for both pools. For example \cite{vanberkum2017bayesian} consider the CMI sub-population in United Kingdom, which consists of assured male lives and is roughly 20\% of the total population, weighted towards pre-retirement ages. \cite{hyndman2013coherent} consider the joint modeling of Female and Male log-mortality by building functional time-series models for their sum and difference (the product-ratio method). This setup is however not applicable to our case, where the pension fund is much smaller and moreover we work with the fixed national mortality table. A related approach to sub-national populations is to first compute the principal Age-components of mortality and then estimate the respective time-dependent coefficients: $m_{x,t} = \sum_i a_i(t) Y^{ref}_i(x) + \epsilon_{x,t}$ \citep{alexander2017flexible}. Gravity models \citep{dowd2011gravity}  work in the Lee-Carter paradigm and introduce dependence of the small population period effect on the larger population:
$$ \kappa_t = \kappa_{t-1} + \phi (\kappa^{ref}_{t-1}-\kappa_{t-1}) + \mu +  \rho \epsilon^{ref}_{t} + \sqrt{1-\rho^2} \epsilon_t,$$ similarly for the cohort term $\gamma_{x-t}$.

A nonparametric framework for capturing the co-dependence of mortality across multiple populations is via Gaussian Processes described in Section \ref{sec:GP}. \cite{huynh2021multi} proposed the use of multi-output GPs to coherently model multiple populations. As an extension, \cite{huynh2021joint} developed hierarchical multi-output GPs to capture multiple cause-of-death mortality rates across different countries and both genders. In these extensions, different populations (or causes of death) translate into distinct levels of a factor covariate, with the latter contributing a cross-population covariance term.  Brazilian national and insurance industry mortality rates are already calibrated or smoothed out through statistical procedures applied by the national institute of statistics (IBGE) or by the association of Brazilian insurers, respectively. Both are mortality tables, so there would be no reason to smooth them again through a (Multi-Output) GP. Therefore, the GP regression will be applied in a single-output approach for the pension fund mortality rates, whether directly or through reference populations.

\subsection{GP modeling} \label{sec:GP}

Gaussian process (GP) models \citep{oakley2002bayesian} are one of the most popular regression methods thanks to the great flexibility they offer in the representation of complex non-linear input-output relationships. GPs define probabilistic models of functional behavior, offering  high prediction power, interpretability, and ability to provide both an interpolation of the data and an uncertainty quantification in the unexplored regions \citep{marrel2008efficient}. The GP model was first developed in spatial statistics under the name of kriging \citep{cressie1990origins}  and rapidly gained attention in the machine learning community. The core idea is to assume the regression function is distributed according to a Gaussian process, which allows to treat the regression function values as unknown quantities and estimate them from the training data. Over the years, the GP model has become popular in a wide range of applications, such as wind power generation \citep{mori2008application}, vehicle design and navigation \citep{Chen2014}, modeling and prediction of natural hazards \citep{rohmer2012meta, liu2017dimension}, etc.

We call a stochastic process $f$ a Gaussian process if for any finite collection of index points $\mathbf{y} = \{y_j, j=1,\ldots,n\}$, the $n$-dimensional multivariate density function of the random vector $\mathbf{f} = [f(y_1), \ldots, f(y_n)]^\top$ is multivariate Gaussian \citep{Rasmussen2005}. A GP is completely specified by its mean and covariance functions. The mean function $m(y) := \mathbb{E}[{f(y)}]$ describes what is the expected value of $f(y)$. The covariance or kernel function, denoted as $c(\cdot,\cdot)$, expresses the degree of dependence between two different function values as a function of two index points, $\Cov(f(y), f(y')) = c(y, y')$. In compact notation, this is  written as
\begin{align*}
f\sim \GP \bigl(m(\cdot), c(\cdot, \cdot) \bigr).
\end{align*}

In the GP regression framework, we usually work with a finite collection of index points which is equivalent to working with a multivariate Gaussian distribution. The collection of index points ${y_j}$, for $j = 1,\ldots,n$, in the previous definition play the role of covariates. In our case, these are age and/or calendar year. The vector of function values whose components are now associated to each of those covariates is then distributed according to a $n$-dimensional multivariate Gaussian distribution,
\begin{eqnarray*}
\begin{bmatrix}
f(y_1) \\
. \\
. \\
f(y_n)
\end{bmatrix} 
\sim {\cal N}\left(
\begin{bmatrix}
m(y_1) \\
. \\
. \\
m(y_n)
\end{bmatrix}, \begin{bmatrix}
c(y_1,y_1) & \cdots & c(y_1,y_n) \\
. & \cdots & .\\
. & \ddots & . \\
c(y_n,y_1) & ... & c(y_n,y_n)
\end{bmatrix} \right)  = {\cal N}( m(\mathbf{y}), \mathbf{C}).
\end{eqnarray*}

For mortality modeling, GPs were introduced in \cite{LUDKOVSKI2016GAUSSIAN} as a data-driven approach for determining age-time dependence in mortality rates and jointly smoothing raw rates across these dimensions. The kernel $c(\cdot,\cdot)$ above then captures the correlation between different covariate coordinates. For example, it is expected that the mortality for age 70 in 2018 or $y_j = (70; 2018)$, is more correlated with $y_p = (69; 2017)$ than with $y_q = (60; 2015)$. 
The GP model provides 
uncertainty quantification associated with smoothed historical experience and generates full stochastic trajectories for out-of-sample forecasts. 
In this paper we analyze univariate GP models that work either with age $y_{ag}$ or with year $y_{yr}$ as the sole covariate, as well as  a bivariate GP model that takes in both $(y_{ag}, y_{yr})$, see Sections \ref{sec:age-dependent}-\ref{sec:gp-models}.

\section{Pension fund population and reference mortality tables} \label{sec:data}

Our analysis is motivated by a dataset summarizing mortality experience of two medium-size pension funds in Brazil that manage pension plans for employees of two firms, which in turn sponsor the pension plan, matching contributions for each one of its workers. By Brazilian law, pension funds are not allowed to write business for non-employees. Our sample is from firms with mainly office workers, belonging to the medium-high income segment. The pension fund population data comprises exposure to risk and number of deaths from 2013 to 2021 (9 years) for male and female employees and pensioners. Approximately, two thirds of the pensioners are males for both samples. Some of the pensioners, especially among the females, are surviving spouses of deceased employees.

\textit{Remark:} In our dataset, the number of observed Female lives and deaths was a fraction of Males, leading to highly sparse exposure and mortality observations, such as majority of ages with zero Female deaths. The resulting level of sparsity rendered the Female data unsuitable for robust model estimation and validation. Therefore, in the analysis below we focus exclusively on Male data. We emphasize that while our methodology is tailored to accommodate small populations (exposures on the order of $[10^3,10^5]$), it nevertheless assumes a minimum data richness to enable smoothing and regularization techniques and capture meaningful patterns across Ages and Years.

\begin{figure}[!ht]
\caption{Time series of total exposure $E_{t} := \sum_x E_{x,t}$ (blue solid lines) and number of deaths $d_{t} := \sum_x d_{x,t}$ (red dashed lines) per year $t$ for both pension funds. Data from 2013 to 2021, males, ages 60--89, pension fund 1 (squares) and pension fund 2 (triangles).}
\centering
\includegraphics[width=0.8\textwidth,trim=0.5cm 1cm 0.5cm 1cm,clip=TRUE]{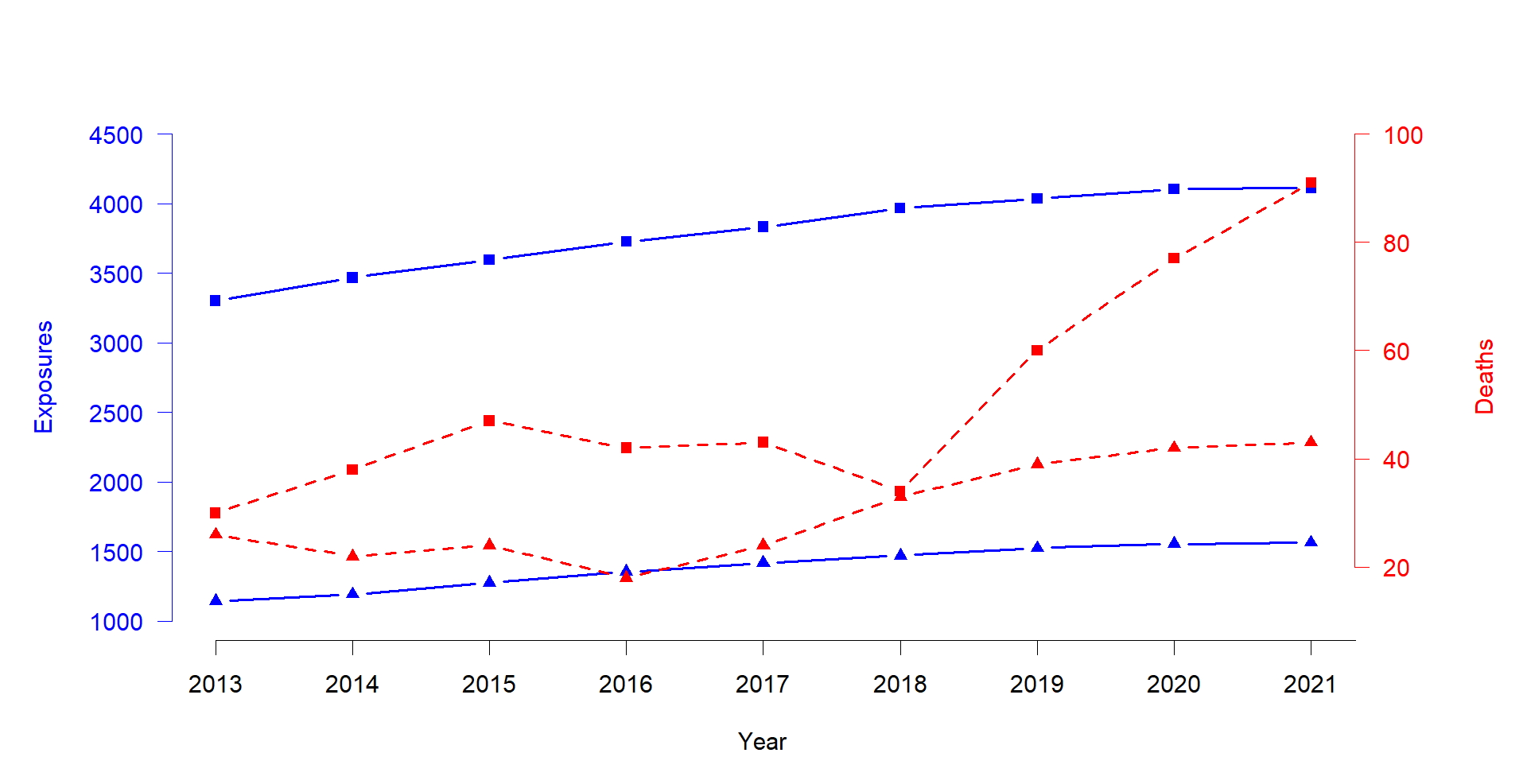}
\label{fig:funds_exp}
\end{figure}

Figure \ref{fig:funds_exp}  shows the annual time series of total exposure $\sum_x E_{x,t}$ to risk (i.e.~number of males pensioners aggregated across ages 60--89) and the annual aggregate number of males' deaths $\sum_x d_{x,t}$.  There is a marked increase in the number of deaths in 2020 and 2021, probably linked to the COVID-19 pandemic.

\begin{figure}[!ht]
\caption{Age pyramids of exposures $E_{x,t}$ (\emph{left}) and number of deaths  (\emph{right}) $d_{x,t}$ per age $x$ for year $t=2018$ for the pension funds 1 and 2: ages 60 to 89.}
\centering
\includegraphics[width=0.8\textwidth]{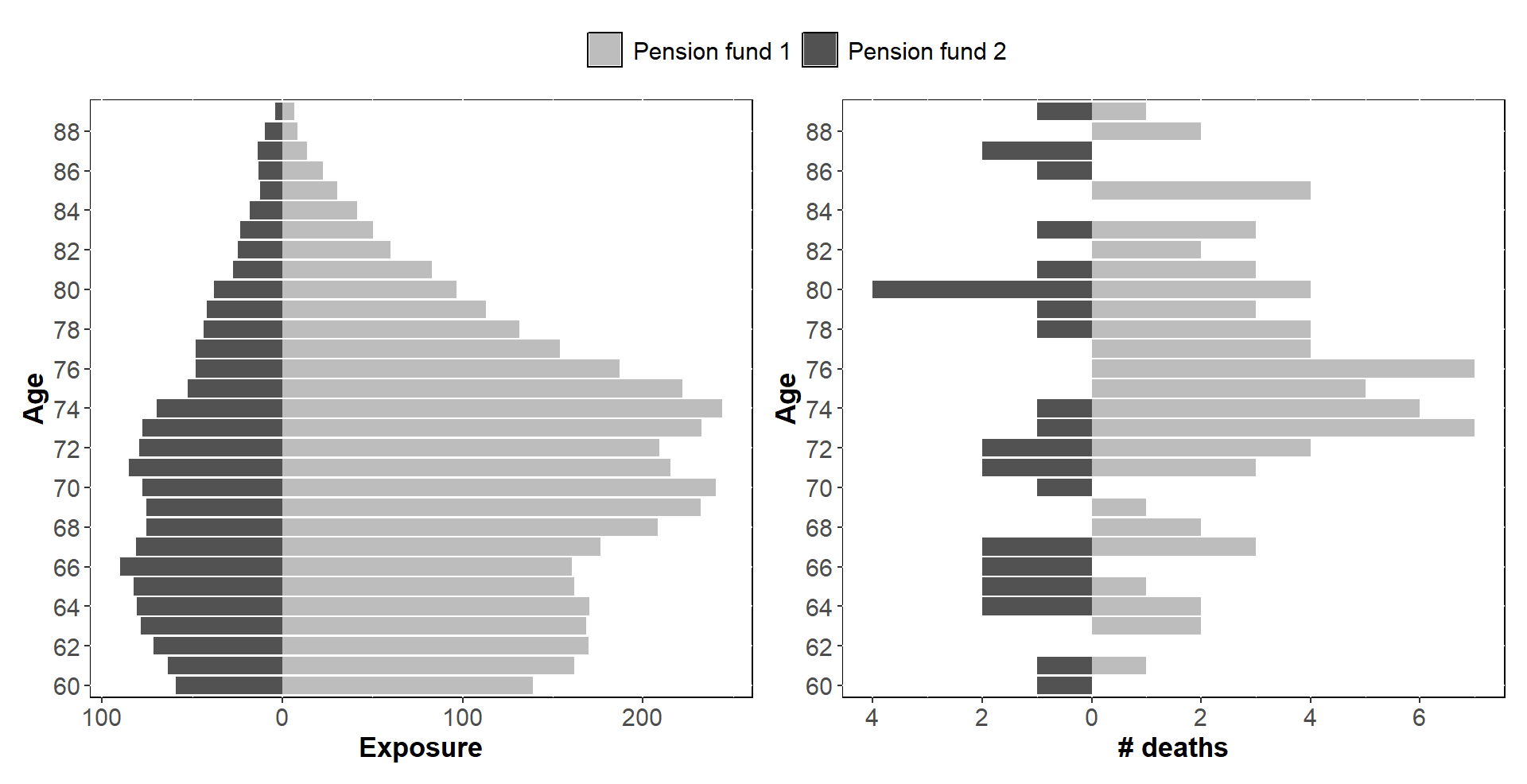}
\label{fig:funds_exp_pyr}
\end{figure}

Figure \ref{fig:funds_exp_pyr} shows the age distribution of pensioners and deaths in the year 2018 for both pension funds when there were $4,113$ and $1,564$ males pensioners on ages 60--89, respectively. For the primary sample, the exposure $E_{x,t}$ varies from 5 (for ages close to 90) to 250 (for ages close to 70). The number of deaths $d_{x,t}$ ranges from none to 7 in 2018. 

\begin{figure}[!ht]
\caption{Log-mortality rates for Males for years 2010/2015/2021, ages 60--80. \emph{Left panel:} Brazilian national population (BRA). \emph{Right}: Brazilian insurance industry table (IND).}
\centering
\includegraphics[width=0.8\textwidth]{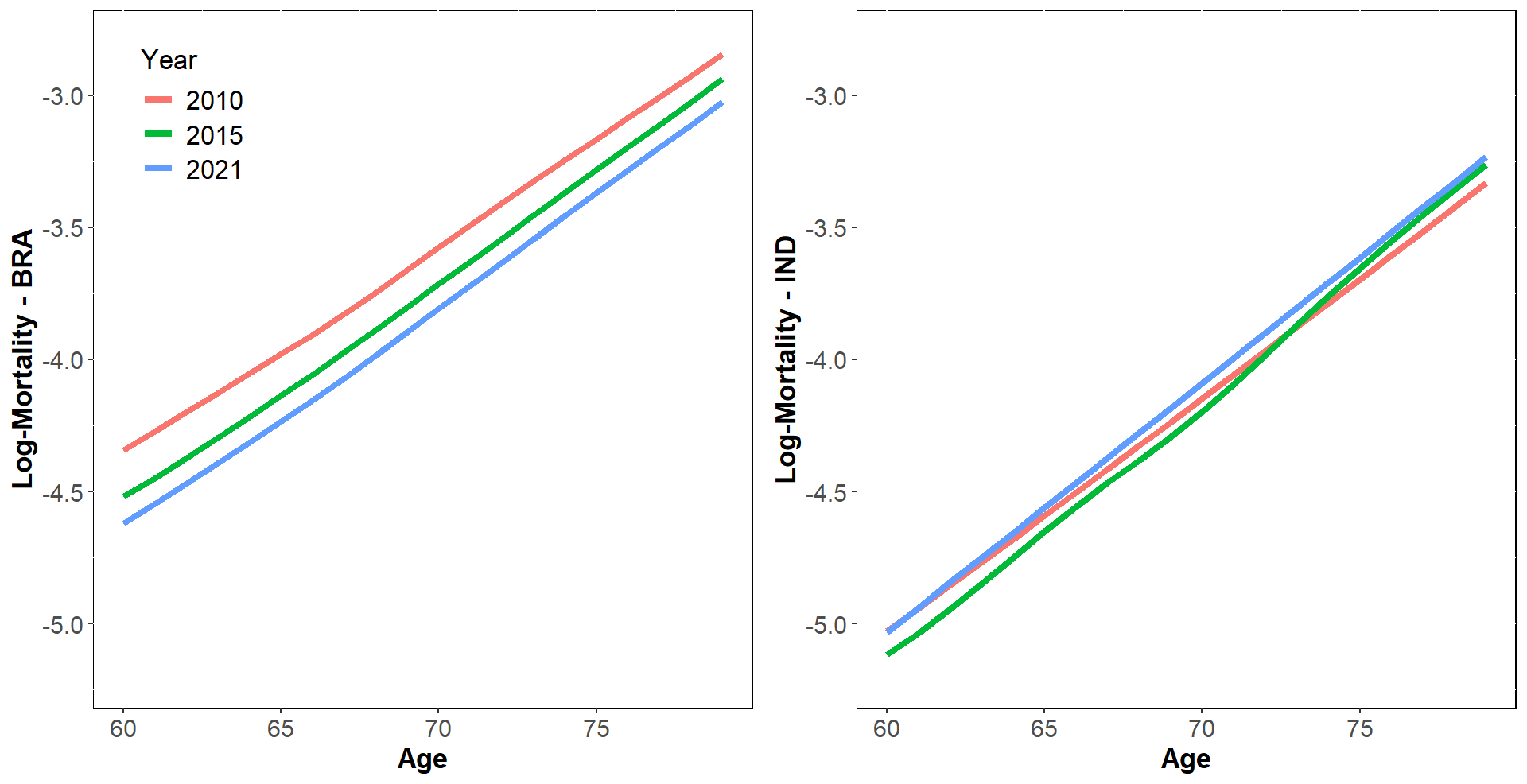}
\label{fig:ibge}
\end{figure}

Figure \ref{fig:ibge} shows the reference Brazilian mortality tables BRA and IND. 
National population mortality tables are published by the Brazilian Institute of Geography and Statistics \citep{ibge2022tables}\footnote{ \url{https://www.ibge.gov.br/estatisticas/sociais/populacao/9126-tabuas-completas-de-mortalidade.html?=t=downloads}}. Data is available from 1998 to 2021, for ages 0 to 80, male, female and both genders. 

It is essential to keep in mind that these tables are heavily post-processed and do not represent raw data. According to \cite{ibge2016tabua}, the Gompertz law (mortality being exponential in age) is used to calibrate mortality rates for ages 60-80. As we are interested in older ages for the pension fund, we used the same IBGE estimated coefficients to extrapolate the mortality rates from age 81 to 89 for each year. In terms of time, the IBGE mortality tables are built based on the results of the Brazilian census which is normally held every 10 years. The 2020 Brazilian Census was postponed to 2022 due to the COVID-19 pandemic.

The insurance industry mortality tables (called BR-EMS) are sponsored by the association of Brazilian insurers (CNSeg) and are based on aggregated data from a subset of the life insurance and open pension plan customers. The tables are  publicly available for download\footnote{\url{http://www.susep.gov.br/setores-susep/cgpro/copep/Tabuas\%20BR-EMS\%202010\%202015\%202021-010721.xlsx}} and are dated as of 2010, 2015 and 2021, see the right panel of Figure \ref{fig:ibge}. The latest 2021 BR-EMS table is based on data up to 2017 and purposely excludes the pandemic period. Differently from the BRA data, the IND mortality rates are not monotonically decreasing across the calendar years for ages $60+$. This is partly due to the fact that there were changes in data quality and collection protocols along the years. In 2021, IND mortality rates are 23\% lower on average than BRA for males at ages 60--89.

Some of the models tested in this paper require a complete reference dataset with mortality tables for each one of the years in the sample period. Therefore, in order to have a smooth interpolation of the IND mortality tables for every year from 2010 to 2021, we fitted a GP model, where the mean function $m(x,t)$ is given by age and calendar year as covariates, cf.\ Section \ref{sec:gp-models}, the kernel is squared-exponential as defined in equation \ref{eq:SqExp-2d}, and the hyperparameters were estimated by maximum likelihood using genetic optimization algorithm \citep{Kriging}. The BRA national mortality tables are available annually, according to an interpolation method described in \cite{ibge2024tables} which is based on a coherent Lee-Carter model, see \cite{unitednations2015}. 

\section{Mortality models}\label{sec:models}

In this section, we introduce the approaches used to model the mortality of the pension funds' populations that were described in Section \ref{sec:data}. First, we introduce the simplest frameworks: the use of a predetermined mortality table with no deflator and the use of a constant deflator relative to a reference population. Then we discuss how one may add age or calendar year dependence on the deflators. We conclude the presentation of the models by introducing GP-based models. All stochastic models in this paper follow the Bayesian paradigm, implying that all parameters should be assigned prior distributions and inference should be made based on posterior distributions. 
Prior distributions are generally vaguely informative, but as all the hyperparameters are easily interpretable, it is expected for specialists to have some degree of expert opinion on the range of the hyperparameters.

\subsection{Practitioners' approach}

The first two frameworks are common for practitioners in the insurance or pension fund industries: the use of known mortality tables with no deflator, or deflated by a single fixed value for all ages.  In the latter case, the fixed effect deflator model may be represented by a simple Generalized Linear Model (GLM) with a Poisson (or Negative Binomial) likelihood, where the offset variable is the product of the reference population mortality rates (IBGE or BR-EMS) and the pension fund exposure. This approach is known in the industry as a \emph{loading deduction} of a determined mortality/annuity table.

These two Fixed Deflator (\textbf{FD}) models are formally defined as:

\ 

Model FD-0 (no deflator). The number of deaths is deterministic given $E_{x,t}$: 
\begin{align} \tag{FD-0} 
d_{x,t} = m^{i}_{x,t}  \cdot E_{x,t}, \qquad\qquad i \in \{BRA, IND\}.
\end{align}

\ \ 

Model FD-1 (constant deflator): 
\begin{align} \tag{FD-1} 
d_{x,t} \mid \theta^i, \omega^i &\sim \NB( e^{\theta^i} \cdot m^{i}_{x,t} \cdot E_{x,t}, \,\omega^i), \qquad i \in \{BRA, IND\} \\
\theta^i &\sim \cN(-0.5, 0.5^2) \nonumber \\
\omega^i &\sim \cN(0, 1^2)I_{(0,\infty)}, \nonumber 
\end{align}
where
\begin{itemize}
    \item $\NB(\cdot, \omega)$ is a Negative Binomial distribution. The notation $d \sim \NB(\mu, \omega)$ corresponds to $\EE[d]=\mu$ and $\Var(d) = \mu (1 + \omega)$. When $\omega = 0$ we recover the Poisson distribution $d \sim \Po(\mu)$; otherwise $\omega$ captures the amount of variance overdispersion compared to the Poisson base case. 
    \item $\theta^i$ is the constant deflator across all ages relative to the reference population $i$. For example, if $\theta^i = -0.5$ then, on average, the mortality of the pension fund is $\exp(-0.5) = 61\%$ of the mortality of the reference population.
\end{itemize}

Note that the ``model'' FD-0 does not have any unknown parameters and therefore no prior distribution is needed. Moreover, it is not stochastic either and it can be retrieved from the model FD-1 when $\theta^i \equiv 0$. For the bonified stochastic model FD-1, for a fixed reference population $i$, both $\theta^i$ and $\omega^i$ are unknown with priors given as above. The need to generalize from the Poisson to the Negative Binomial observation likelihood is driven by the well known overdispersion of death counts, cf.~Section \ref{sec:overdispersion} below.

\subsection{Age-dependent deflators} \label{sec:age-dependent}

In this subsection we present the age-dependent deflators' models (\textbf{AD}). All three models presented share the same likelihood function, their difference being in the structure (or lack thereof) imposed on the deflators.

The age-dependent deflator fixed effects model (AD-FE) does not impose any structure on the deflators, allowing, in principle, completely unrelated deflators across ages. The model is formally defined as follows:

Model AD-FE (Age-dependent deflators fixed effects):
\begin{align} \tag{AD-FE} 
d_{x,t} &\sim \NB \left( e^{\theta^i_{x}} \cdot m^{i}_{x,t} \cdot E_{x,t}, \, \omega^i \right), \qquad i \in \{BRA, IND\} \\
\theta^i_{x} &\sim \cN(-0.5, 0.5^2) \nonumber \\
\omega^i &\sim \cN(0, 1^2)I_{(0,\infty)}, \nonumber 
\end{align}
where $\theta^i_{x}$ is the deflator at age $x$ relative to the reference population $i$. The priors are assigned independently, see Table \ref{tab:modSOGP}. Note that FD-1 is a special case of AD-FE with constant $\theta^i_x$ across ages.

In order to impose structure on age deflators, an autoregressive age-dependence among $\theta^{i}_{x}$'s has been suggested. In existing literature, \cite{vanberkum2017bayesian} proposed an autoregressive structure on $\theta^i_{x}$; \cite{li2017coherent} also uses an autoregressive approach to the age-deflators which is then applied to a reference population; \cite{oliveira2021bayesian} uses a random walk for $\theta^i_{x}$ in terms of $x$. 

\ \ 

Model AD-AR (Age-dependent deflators with autoregressive predictor):
\begin{align}\tag{AD-AR}
d_{x,t} \mid \theta^i_{x}, \omega^i &\sim \NB \left( e^{\theta^i_{x}} \cdot m^{i}_{x,t}  \cdot E_{x,t}, \, \omega^i \right)  \qquad i \in \{BRA, IND\} \\
\theta^i_{x} \mid \theta^i_{x-1}, \rho^i &\sim \cN \left( \mu^i + \rho^i \theta^i_{x-1}, 0.5^2(1-(\rho^i)^2) \right), \nonumber \\
\theta^i_{60} &\sim \cN(-0.5, 0.5^2 ) \nonumber \\
\omega^i &\sim \cN(0, 1^2)I_{(0,\infty)} \nonumber \\
\rho^i &\sim \cN(1, 1^2)I_{(0,1)} \nonumber \\
\mu^i &= -0.5 \times (1 - \rho^i) \nonumber.
\end{align}

The structure of the AD-AR model is similar to the one used in \cite{vanberkum2017bayesian}, where the prior choices ensure that the unconditional mean and variance of the deflators are the same for all ages. To see that, we first compute the mean and the variance of the deflators conditional on the autoregressive parameter $\rho^i$ (which is itself random). As the prior for $\rho^i$ restricts its values to the interval $(0,1)$, the auto-regressive process for $\theta^i_{x}$ is stationary and therefore
$$\EE[ \theta^i_{x} \mid \rho^i] = \frac{\mu^i}{1- \rho^i} = -0.5 \text{ and } \text{Var}(\theta^i_{x} \mid \rho^i) = \frac{0.5^2(1-(\rho^i)^2)}{1-(\rho^i)^2} = 0.5^2.$$
Then, using the Law of Total Expectation/Variance, we observe that
\[\EE[ \theta^i_{x}] = \EE[ \EE[ \theta^i_{x} \mid \rho^i] ] = -0.5 \]
and
\[ \Var(\theta^i_{x}) = \EE[\Var(\theta^i_{x} \mid \rho^i)] + \Var(\EE[\theta^i_{x} \mid \rho^i]) = \EE[0.5] + 0 = 0.5.\]

As $\EE[\theta^i_{60}] = -0.5$ and $\text{Var}(\theta^i_{60}) = 0.5$, our prior assigned for the deflators in the AD-AR model is consistent with those from the AD-FE model.

In both models, for a reference population, the deflators for one age depend (stochastically) on the deflator of the previous age. In the AD-AR model the parameter $\mu^i$ can be interpreted as the base level for all the deflators, regardless of age, while $\rho^i$ provides the strength of the dependence between any age and its preceding one. As the deflator of each age depends on the deflator of the immediately younger group, we also set a probability distribution for the first age group being analyzed, i.e., the 60 year olds.

As a different generalization of AD-FE, we consider GP models for the age-dependent deflators (AD-GP) relative to the reference population. Instead of forcing an Auto-Regressive structure across age groups, the AD-GP model postulates that the dependence between deflators for different ages is given by a Gaussian Process driven by $y_{ag}$. We use the squared-exponential kernel that takes the form
\begin{eqnarray}\label{eq:SqExp-1d-age}
c_{ag}(y_j,y_p) := \sigma^2 \exp\left( - \frac{(y_{ag}^j - y_{ag}^p )^2}{2\phi_{ag}^2} \right).
\end{eqnarray}
This kernel has two hyperparameters: the process variance $\sigma^2$ and the lengthscale $\phi_{ag}$.

For reference population $i$, let $x \mapsto \theta^i(x)$ denote now a function representing the deflator for age $x$. To be consistent with the previous models, when $x \in \{60, 61, \ldots, 89\}$, we write $\theta^i(x) = \theta_{x}^i$, but the reader should note that $\theta^i(\cdot)$ can be applied to any $x$, even if it is not an integer value. In the AD-GP model this function is assumed a priori to follow a GP.

\ \ 

Model AD-GP (Age-dependent deflators with Gaussian Process):
\begin{align*} \tag{AD-GP}
        d_{x,t} \mid \theta^i(\cdot), \omega^i &\sim \NB(e^{\theta^i({x})} \cdot m^{i}_{x,t}  \cdot E_{x,t}, \, \omega^i) \\
    \theta^i(\cdot) \mid \sigma^i, \phi^i_{ag} &\sim \GP( -0.5, c_{ag}^i(\cdot, \cdot)) \\
    \omega^i &\sim \cN(0,1^2)I_{(0,\infty)} \\
    (\sigma^i)^2 &\sim \cN(0.5, 0.5^2)I_{(0,\infty)} \\
    \phi_{ag}^i &\sim \cN(4,4^2)I_{(0,\infty)}.
\end{align*}

As an abuse of notation, the mean function in the second row above is denoted as a single number, $-0.5$, meaning the deflator for all ages has the same expected value {\it a priori}.

To interpret the structure of \eqref{eq:SqExp-1d-age}, note that the covariance reaches its maximum value $c_{ag}(y_j,y_p) = (\sigma^i)^2$ when $y^j =y^p$, and when the ages $y^j$ and $y^p$ are far apart $c_{ag}(y^j; y^p) \approx 0$, so that $\theta^i(y^j)$ and $\theta^i(y^p)$ are statistically independent. The lengthscale parameter $\phi^i_{ag}$ controls how fast the dependence between two deflators decays in age. If $\phi^i_{ag}$ is large, the dependency between two different values of the function decays very slowly and the function $\theta^i$ is expected to vary only a little, resembling almost a linear function. The process variance parameter $\sigma^2$ controls the magnitude of variation (i.e.,~the amplitude) of $\theta^i(\cdot)$. 

For a fixed age $x$, we have that, a priori,
\[ \theta^i({x}) \mid  \sigma^i, \phi^i_{ag} \sim \cN \left(\beta^i_{0}, (\sigma^i)^2 \right). \]

Therefore,
\[\EE[\theta^i(x)]  = \EE[ \EE[\theta^i_{x} \mid \sigma^i, \phi^i_{ag} ]] =  \EE[-0.5 ] = -0.5 \] 
and
\begin{align*}
\text{Var}(\theta^i(x)) &= \text{Var}(\theta^i_{x}) 
= \EE[ \text{Var}(\theta^i_{x} \mid \sigma^i, \phi^i_{ag} )] + \text{Var}(\EE[\theta^i_{x} \mid \sigma^i, \phi^i_{ag} ]) \\
    &=  \EE[ (\sigma^i)^2] + \text{Var}(-0.5) 
    = 0.5 + 0.
\end{align*}

\emph{Remark:} In AD-GP, the deflator function $\theta^i(\cdot)$ is taken to have the SE kernel $c_{ag}$  \eqref{eq:SqExp-1d-age} in age. In fact, the AD-AR model can also be viewed as a GP model with the exponential Mat\'ern-1/2 kernel $\theta^i(\cdot) \sim GP(-0.5, c^i_{M12,ag}(\cdot, \cdot))$, with $c_{M12,ag}(y^j, y^p) = \exp( - \phi| y^j_{ag} - y^p_{ag})| )$ for a (distinct from $\phi_{ag}$) lengthscale hyperparameter $\phi \in \mathbb{R}_+$.

\subsection{Time-dependent deflators}

In this subsection we present the time-dependent deflators' models (TD), which follow the same structure as the age-dependent ones from Section \ref{sec:age-dependent}. Instead of focusing on the role of age, our interest lies in introducing time-dependence for the deflators to capture the idea that the evolution of the pension funds' mortality and of the national populations follow different trends. 

Similar to the notation in Section \ref{sec:age-dependent}, we denote by $\theta^i(t)$ the function that returns the deflator for the year $t$ when the reference population $i$ are fixed. We first consider the autoregressive and the random walk specifications:

\ \ 

Model TD-AR (time-dependent deflator with autoregressive predictor):
\begin{align}\tag{TD-AR}
d_{x,t} \mid \theta^i(\cdot), \omega^i &\sim \NB( e^{\theta^i(t)} \cdot m^{i}_{x,t}  \cdot E_{x,t}, \, \omega^i)\\
\theta^i(t) \mid \theta^i(t-1), \rho^i &\sim \cN \left( \mu^i + \rho^i \theta^i(t-1), 0.5^2(1-(\rho^i)^2) \right), \nonumber \\
\theta^i(2013) &\sim \cN(-0.5, 0.5^2 ) \nonumber \\
\omega^i &\sim \cN(0, 1^2)I_{(0,\infty)} \nonumber \\
\rho^i &\sim \cN(1, 1^2)I_{(0,1)} \nonumber \\
\mu^i &= -0.5 . (1 - \rho^i). \nonumber
\end{align}

\noindent Similarly to the AD-AR setup, the TD-AR model defines a parametric stochastic relationship between the mortality deflators of different calendar years. Due to its similarities, the interpretation of the parameters is the same too.

As it may be hard, {\it a priori}, to pinpoint how the deflators in different years relate to each other, we propose the TD-GP model ---the equivalent of AD-GP in the time domain. The analogous calendar-year  kernel is
\begin{eqnarray}\label{eq:SqExp-1d-year}
c_{yr}(y_j,y_p) := \sigma^2 \exp\left( - \frac{(y_{yr}^j - y_{yr}^p )^2}{2\phi_{yr}^2} \right).
\end{eqnarray}

\ \ 

Model TD-GP (time-dependent deflator with GP structure):

\begin{align*} \tag{TD-GP}
        d_{x,t} \mid \theta^i(\cdot), \omega^i &\sim \NB(e^{\theta^i(t)} \cdot m^{i}_{x,t}  \cdot E_{x,t}, \, \omega^i) \\
    \theta^i(\cdot) \mid \sigma^i, \phi^i_{yr} &\sim \GP( -0.5, c_{yr}^i(\cdot, \cdot)) \\
    \omega^i &\sim \cN(0,1^2)I_{(0,\infty)} \\
    (\sigma^i)^2 &\sim \cN(0.5, 0.5^2)I_{(0,\infty)} \\
    \phi_{yr}^i &\sim \cN(4,4^2)I_{(0,\infty)}.
\end{align*}
Once again, the GP kernel is chosen from the squared-exponential family \eqref{eq:SqExp-1d-year} and the  GP prior mean is constant across different years.

\subsection{Direct modeling of the pension fund} \label{sec:gp-models}

The last set of models we consider employ Gaussian Processes to model the pension fund population by itself, with no reference population. In lieu of having a baseline mortality, we de-trend the data using parametric prior means. Both a univariate GP that captures only Age dependence, as well as a bivariate Age-Year GP are considered. The respective prior means are: (i) a \cite{gompertz1825xxiv} curve with no calendar year covariate (model GP-S1), and (ii) a Gompertz curve with a calendar year covariate (model GP-S2). 

Model GP-S1 (S for single-population) directly captures the pension fund log-mortality $\psi(x)$, postulating it to be a function of age $x$ and employing the GP to map from $x$ to $\psi({x})$. GP-S1 assumes that the log-mortality is independent of time $t$ and uses a prior mean
$\mu_{age}(x) = \beta_{0} + \beta_{ag} x$. This linear trend in $x$ for $\psi(\cdot)$ matches the Gompertz assumption of mortality growing exponentially in age. As with all GP models in this paper, the kernel is chosen as the squared-exponential family specified in \eqref{eq:SqExp-1d-age}.

\ \ 

Model GP-S1 (Single Population GP with Gompertz prior):
\begin{align*}\tag{GP-S1}
    d_{x,t} \mid \psi(\cdot), \omega &\sim \NB(e^{\psi(x)} \cdot E_{x,t},  \, \omega) \qquad \\
    \psi(\cdot) \mid \beta_{0}, \beta_{ag}, \sigma^2, \phi &\sim \GP \left( \mu_{ag}(\cdot), c_{ag}(\cdot, \cdot) \right) \\ 
    \mu_{ag}(x) &= \beta_{0} + \beta_{ag} (x-60) \\
    \omega &\sim \cN(0,1^2)I_{(0,\infty)} \\
    \beta_{0} &\sim \cN(-5, 1^2) \\
    \beta_{ag} &\sim \cN(0.1, 0.1^2) \\
    \sigma^2 &\sim \cN(0.5, 0.5^2) I_{(0,\infty)} \\
    \phi_{ag} &\sim \cN(4,4^2)I_{(0,\infty)}.
\end{align*}

We emphasize that $\psi_x$ is now the log-mortality rate, so it is on a different scale compared to the deflators $\theta_x$ considered in the previous section.
To guarantee our prior views are consistent across all models, we compare model GP-S1 with AD-GP. Based on the likelihoods, both mortality rates should be the same, implying that
\[e^{\psi_{x}} E_{x,t} = e^{\theta^i_{x}} m^{i}_{x,t}   E_{x,t} \Rightarrow \psi_{x} = \theta^i_{x} + \log( m^{i}_{x,t}). \]

Conditioning on the parameters of both models, on the one hand we have that
\begin{align}\label{eq:consistency1-gp-s1}
    \EE[ \psi_{x} \mid \ldots] = \beta_{0} + \beta_{ag} (x-60) 
\end{align}
and on the other, to have consistency with AD-GP,
\begin{align}\label{eq:consistency2-gp-s1}
    \EE[ \psi_{x} \mid \ldots] = \EE[ \theta^i_{x} + \log( m^{i}_{x,t}) \mid \ldots].
\end{align}

Therefore, taking the expectation in \eqref{eq:consistency1-gp-s1} and \eqref{eq:consistency2-gp-s1}, we would like to have
\begin{align}\label{eq:consistency3-gp-s1}
    \EE[\beta_{0}] + \EE[\beta_{ag}] (x-60) = -0.5 + \log( m^{i}_{x,t}),
\end{align}
where we used the fact that under model AD-GP assumptions the unconditional mean of the deflators is set to $-0.5$ for all ages. In order to define a reasonable prior mean for $\beta_0$ and $\beta_{ag}$ we do a standard linear regression of $m^i_{x,t}$ against $(x-60)$ (note that $g$ is fixed but $t$ is not, so there are multiple observations for each age $x$) and then set the prior mean of $\beta_0$ to the resulting $y$-intercept minus $0.5$, and the prior mean of $\beta_{ag}$ to the resulting slope in $(x-60)$. For the population under analysis the prior means for $\beta_0$ and $\beta_{ag}$ are then chosen as, respectively, $-5.0$ and $0.1$, when using the reference population as $i=BRA$.

\emph{Remark:} placing a prior mean can be interpreted as parametrically de-trending the original data and then using the non-parametric GP to model the residuals. Because the globally imposed $\beta$'s in $\mu_{ag}$ create correlation between $\psi_{x}$'s as $x$ varies, the latter residual GP should have a weaker dependence structure, i.e., a shorter lengthscale $\phi^i_{x}$ that decorrelates the above residuals relatively quickly across different ages.  

Model GP-S2 generalizes GP-S1 by fitting a two-dimensional GP for the mortality surface in Age-Year. Accordingly, it also employs a bivariate prior mean that accounts for both the age $x$ and the year $t$ in the GP modeling the pension fund's log-mortality. The GP kernel in this case is chosen as the separable bivariate squared exponential kernel 
\begin{eqnarray}\label{eq:SqExp-2d}
c(y_j,y_p) = \sigma^2 \exp\left( - \frac{(y_{ag}^j - y_{ag}^p )^2}{2\phi_{ag}^2} - \frac{(y_{yr}^j - y_{yr}^p )^2}{2\phi_{yr}^2} \right)
\end{eqnarray}
with three hyperparameters $\sigma, \phi_{ag}, \phi_{yr}$.

\medskip 

Model GP-S2 (Single Population bivariate GP with linear prior mean):
\begin{align*}\tag{GP-S2}
    d_{x,t} \sim \NB(e^{\psi({x,t})} \cdot  E_{x,t}, \, \omega) \qquad \\
    \psi(\cdot,\cdot) \mid \beta_{0}, \beta_{ag},  \beta_{yr}, \sigma^2, \phi_{ag}, \phi_{yr}  \sim \GP( \mu_{ag,yr}(\cdot,\cdot), c(\cdot, \cdot)) \qquad \\
    \mu_{ag,yr}(x,t) = \beta_{0} + \beta_{ag} (x-60) + \beta_{yr} (t-2013) \qquad \\
    \beta_{yr} \sim \cN(0,0.1^2) \qquad \\
    \phi_{yr} \sim \cN(4,4^2)I_{(0,\infty)}. \qquad  
\end{align*}
Revisiting \eqref{eq:consistency3-gp-s1} we wish to have
$$
\EE[\beta_{0}] + \EE[\beta_{ag}] x + \EE[ \beta_{yr}] t = -0.5 + \log( m^{i}_{x,t})
$$
which is implemented by doing a multiple linear regression of $\log(m^{i}_{x,t})$ against age and calendar year $(x,t)$ (keeping reference population $i$ fixed) and then matching the respective coefficients in Age, in Year. and the constant term (minus the $-0.5$ offset). After performing this exercise for $i=BRA$, we choose the priors for $\beta_0$, $\beta_{ag}$ and $\beta_{yr}$ to be centered around $-5.0$, $0.1$ and $0$, respectively.

Prior distributions for the hyper-parameters of all models are summarized in Table \ref{tab:modSOGP} at Appendix A.

\section{Fitted Models}\label{sec:results}
In this section we present the results obtained from training the 8 models (FD-1, AD-AR, AD-FE, AD-GP, TD-AR, TD-GP, GP-S1, GP-S2)  on data from 2013 to 2018, utilizing the Brazilian population as the baseline ($i=BRA$). Supplementary analysis for an alternative pension fund presented in Appendix \ref{sec:alternative_pension_fund}.

Our models were implemented in a Bayesian framework using the \texttt{Stan} library in $\texttt{R}$ 
\citep{CARPENTER2017,rstan2018}. We implemented the GP models with a method for scaling inference of GPs with stationary kernels in \texttt{Stan} using the Fast Fourier Transform, as described in \cite{HoffmannOnnela2023}. We run 3 chains, each with $10,000$ samples, a burn-in of $2,000$ and a thinning factor of $20$. This provided $1200$ samples for the posterior and predictive distributions for each model\footnote{GP deflators models run in approximately 30 minutes, whereas GP-S1/2 models take about an hour on a Intel Core i5 CPU @ 1.70GHz and 4 cores.}.

\subsection{Overdispersion parameter} \label{sec:overdispersion}

\begin{figure}[!ht]
\caption{Inferred overdispersion parameters $\omega$ for Males and BRA reference population. We show the prior density (red curve), the posterior histogram and posterior mean (solid vertical line) 90\% quantile range (dashed lines).\label{fig:posterior_omega}}
\centering
\medskip

\begin{tabular}{ccc}
\includegraphics[width=0.3\textwidth]{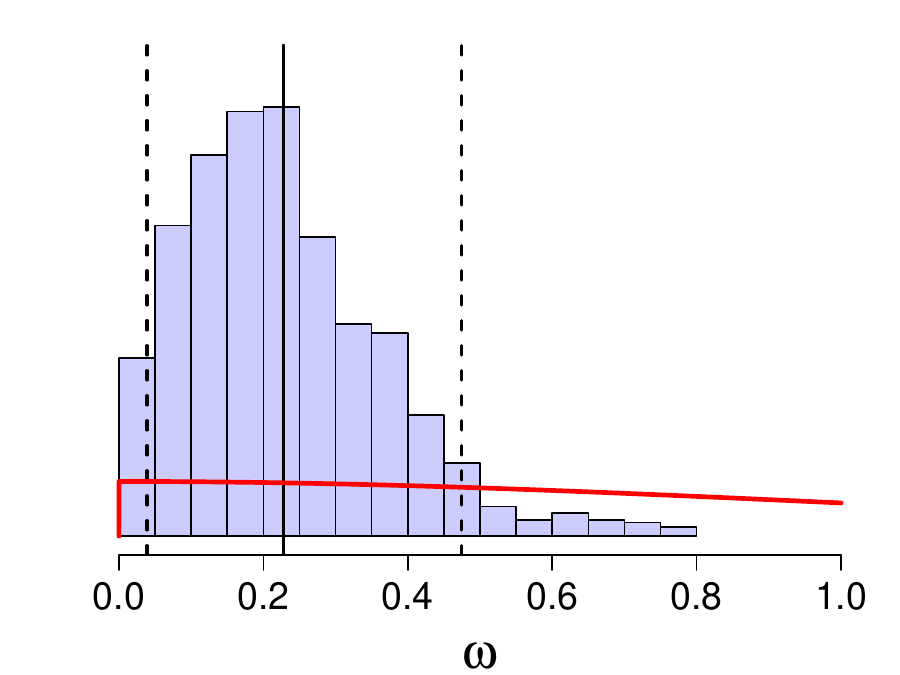} &
\includegraphics[width=0.3\textwidth]{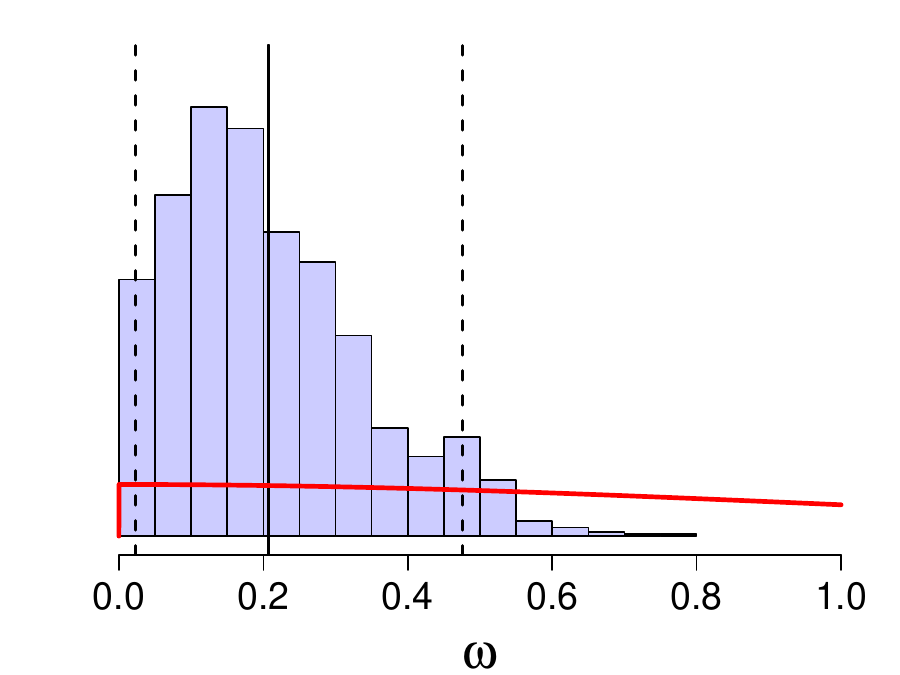} &
\includegraphics[width=0.3\textwidth]{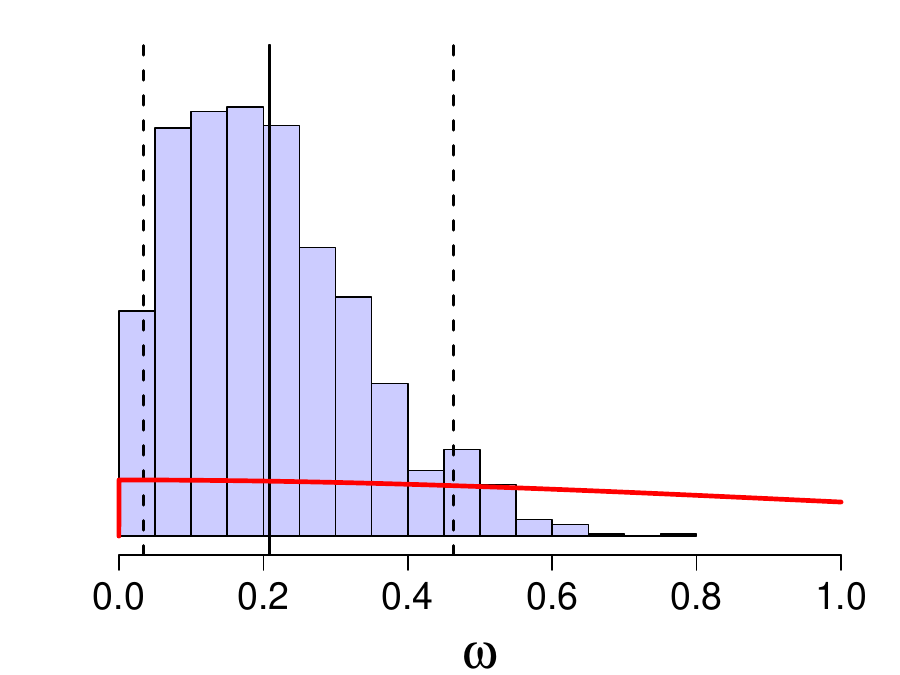} \\
FD-1 & AD-GP & TD-AR \\
\end{tabular}

\end{figure}

Since all the eight models presented are based on a Negative Binomial likelihood, the estimated overdispersion parameter $\omega$ is comparable across models. The histograms of the posterior distributions of $\omega$ for several of the models is shown in Figure \ref{fig:posterior_omega}. The results are qualitatively similar: the mode of the posterior distribution of $\omega$ shifts to the right compared to the prior. For most models, the posterior mode is around $0.2$, implying that the conditioned variance of the observed number of deaths is $20\%$ higher than its conditioned mean, confirming the expected overdispersion that has been repeatedly observed in small mortality samples. The fact that the posterior of $\omega$ is consistent across all models provides internal validation that all the trainings converged and that the assumptions about the deflators are not impacting the assumed relationship between latent mortality and observed deaths. The $90\%$ posterior intervals are far from zero, so that we can reject the hypothesis of $\omega=0$ at $95\%$ confidence level and conclusively infer that the likelihood distribution is not Poisson.

Across all fitted models, we find that estimates of $\omega$ are significantly bigger than zero with high probability. For AD-GP, $\omega^{IND}$ has mean 0.20 $-$ it means that the estimate of variance is 20\% higher than the mean. 90\% credibility interval on $[0.03; 0.47]$, so $\omega^{IND} > 0.03$ with 95\% posterior probability. In particular, the posterior probability of $\omega > 0.05$ exceeds 90\%, even though we always start with a very weakly informative prior distribution (red line). Therefore, $H_0: \omega = 0$ may be rejected.

\subsection{Deflators} \label{sec:deflators}

\begin{figure}[!ht]
\caption{Inferred deflators $\theta^i(\cdot)$ for Males and BRA reference population across six models. For the FD-1 model we show the prior and posterior densities. 
For all other models, thicker error bars denote the 50\% posterior credible interval, thinner bars the 90\% interval, and the dots the posterior mean. For models AD-FE, AD-AR and AD-GP the horizontal axis denotes age, while for TD-AR and TD-GP it denotes year. For the time-dependent models, 2019 is a forecast.}
\label{fig:posterior_theta}
\centering
\medskip

\begin{tabular}{ccc}
\includegraphics[width=0.3\textwidth]{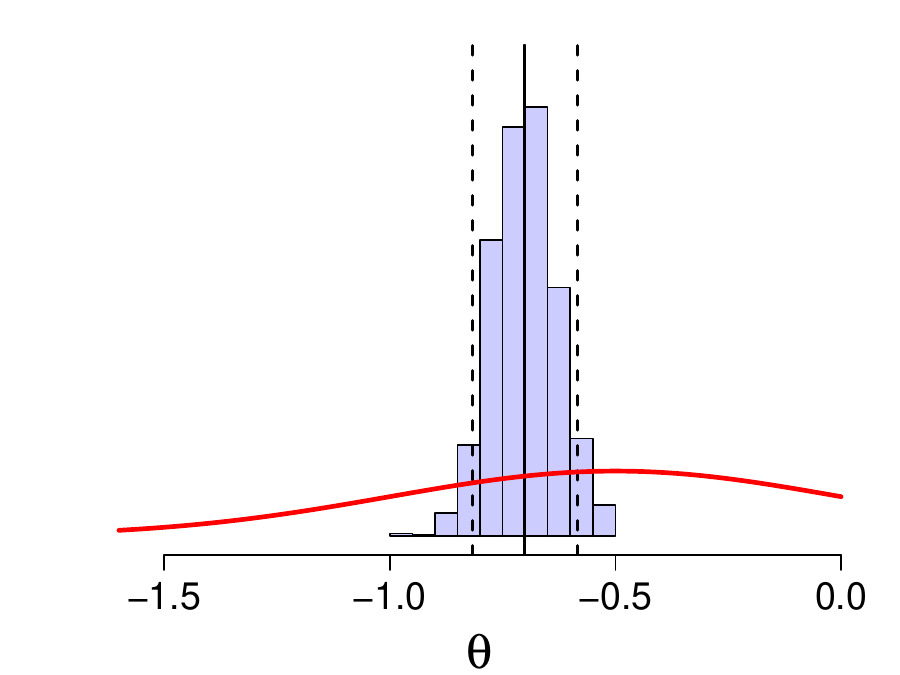}  &
\includegraphics[width=0.3\textwidth]{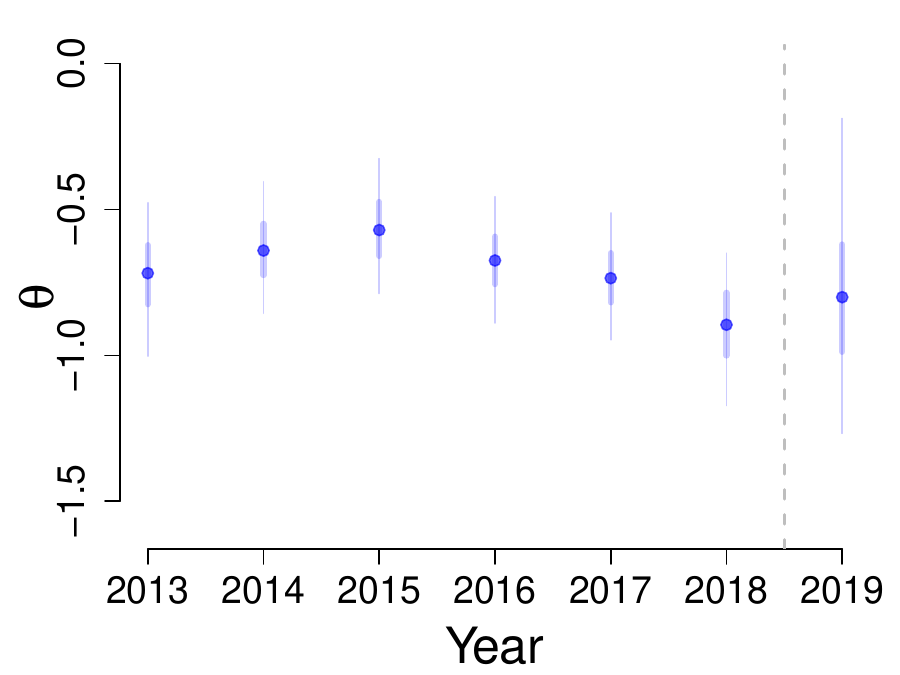} &
\includegraphics[width=0.3\textwidth]{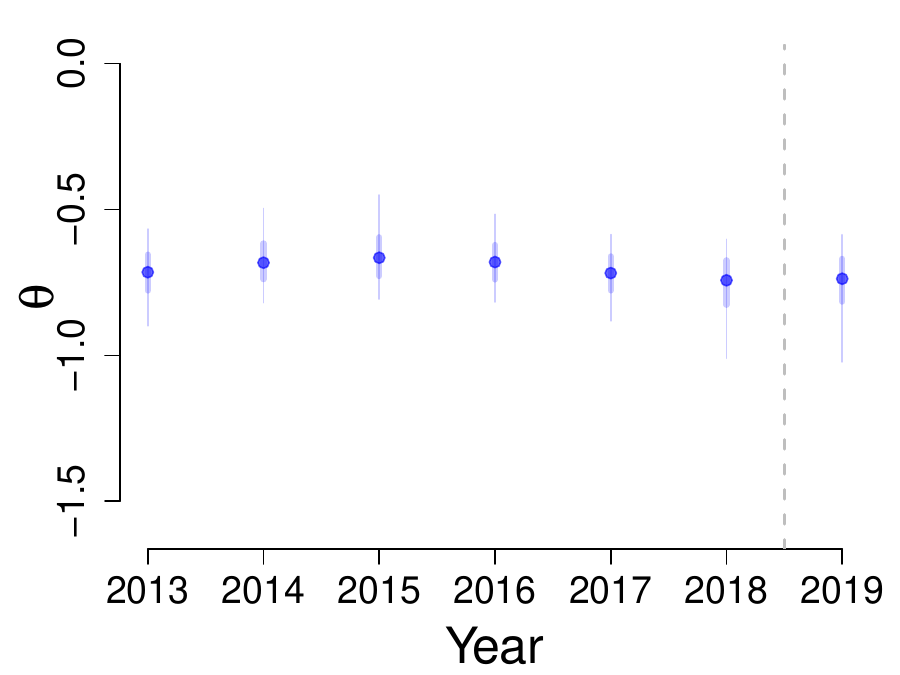} \\
FD-1 & TD-AR & TD-GP \\
\includegraphics[width=0.3\textwidth]{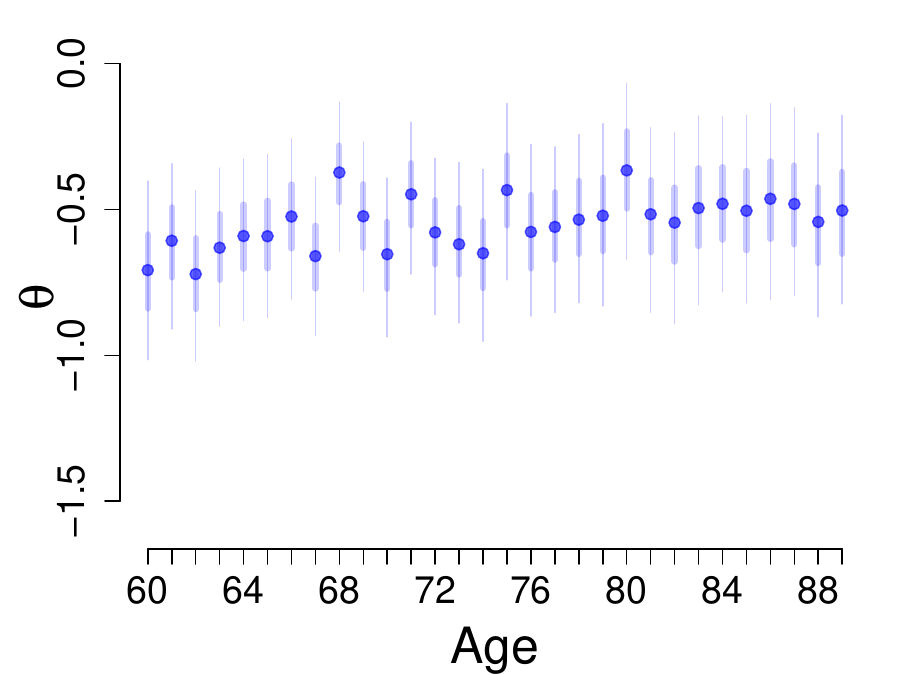} &
\includegraphics[width=0.3\textwidth]{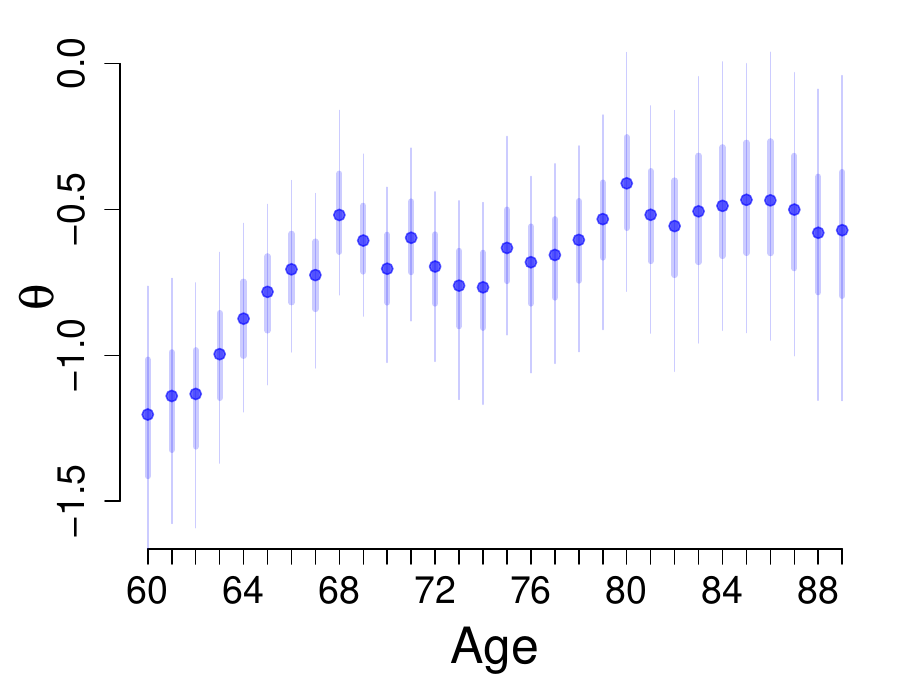} & 
\includegraphics[width=0.3\textwidth]{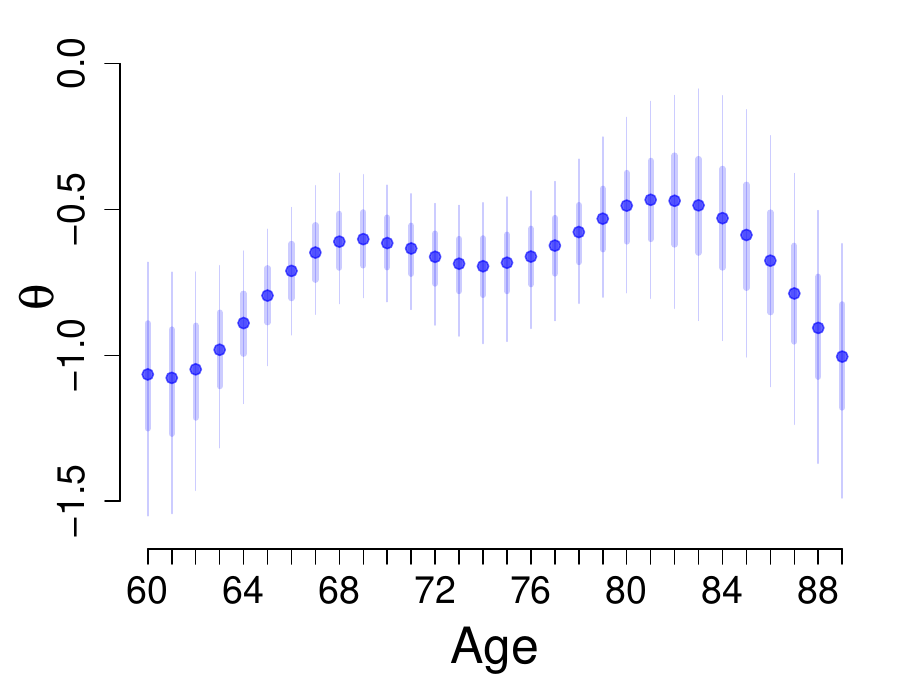} \\
AD-FE & AD-AR & AD-GP \\
\end{tabular}

\end{figure}

Before analyzing specific model parameters, we compare the estimated fund deflators $\theta$ across the models. Recall that apart from model FD-1, we do not directly assign prior distributions to $\theta^i$; rather the prior for $\theta^i$ (and hence its posterior) is induced by priors of other model parameters.
Figure \ref{fig:posterior_theta} presents the posterior distribution of the deflators in each of the models. Model FD-1 differs from all the others, as it only has a single age- and year-independent deflator, whose posterior distribution is presented as a histogram. For the three age-dependent (AD-) models the horizontal axis denotes the different ages, while in the time-dependent models (TD-) it denotes the year.

The single deflator estimated in FD-1 has a posterior density with mass concentrated between $-1.0$ and $-0.5$; a similar range is observed for the estimates in all other models. Hence the fund mortality is in the range of $\exp(-1)=36.8\%$ to $\exp(-0.5)=60.6\%$ of the national mortality.
The time trend is difficult to discern as we have access to just six training years (2013--2018); nevertheless both TD-AR and TD-GP suggest a downward trend. The take-away is that there is a weakly identifiable decrease in $\theta$'s over the years, i.e., the relative gap between the fund mortality and the national population is getting wider. In Figure \ref{fig:posterior_theta} we also present the predictive distribution for the deflators in 2019, which, as expected, yields a wider predictive range. 

For pension fund 1, the age-dependent models highlight that the deflators are strongly non-constant across ages, indicating a major deficiency of FD-1 that assumes a constant $\theta$. The disparity of deflators is more than $\max_{x,x'} |\theta_x - \theta_{x'}| > 0.5$ across ages, translating into 30\% relative ratio change. As we confirm below, a more flexible structure than FD-1 is necessary for acceptable goodness-of-fit. In the specific pension fund, we find a systematic pattern of increasing values of $\theta_x$ for younger ages ($x \in [60,70]$), a U-shape between ages 70-80 and decreasing deflators for ages above $x>80$. The models disagree on the strength of the pattern for ages 80-90. 
Although the deflators are significantly age-dependent, the quantity of the available training date leads to quite wide uncertainty bands of the inferred deflators. As a result, there is insufficient data to identify the fine features of this age-dependence, as highlighted by the differing patterns of AD-AR and AD-GP models.

It is also instructive to study the posterior uncertainty and variability of $\theta_x$ across ages. The haphazard pattern of the AD-FE estimates showcases the need to impose structure, otherwise deflators for adjacent ages might end up drastically different (see ages 68 and 69 in Figure \ref{fig:posterior_theta}). AD-AR achieves some of this smoothing, however it also has very wide posterior uncertainty, indicating that the model is not able to learn much.  Due to the smooth nature of GPs, its estimates are less volatile. This is a desirable feature, as we expect the gap between the funds' mortality rates and the country-wise mortality to be similar for nearby age groups. In turn, the assumed smoothness of the GP deflators allows for significant information fusion across ages which leads to tighter posterior uncertainty bands and hence the most informative take on the age-structure of $\theta$'s. Information fusion is most pronounced at the middle of the training range, note how the AD-GP error bars for ages 68-78 are significantly narrower compared to ages above or below, and much narrower than those same ages in AD-FE. Similarly, the error bars of TD-GP are almost twice as narrow as those of TD-AR. Even more pronounced, the assumed smoothness in year, makes the out-of-sample predictive band of TD-GP for 2019 to be much tighter than that of TD-AR. 

\subsection{Inferred mortality rates and death count distribution}
We next shift our focus to the overall predictions of the pension fund mortality in terms of respective mortality rates and predicted death counts, including uncertainty thereof. Depending on the model choice, mortality of the pensioners is either captured directly, or implied by the corresponding deflator. 

\begin{figure}[!ht]
\caption{\emph{Top:} Predicted pension fund's Male log-mortality rates as a function of age $x$ for year 2019 (blue) induced by models FD-1, AD-GP and GP-S2. We compare to reference population (BRA) mortality curve (green squares) and raw mortality (red crosses). \emph{Bottom:} Predicted number of deaths for each age $x \in \{60, \ldots, 89\}$ in $t=2019$ (blue) relative to the observed number $d_{x,t}$ of deaths (red crosses). Circle sizes represent the probability assigned to the corresponding number of deaths, thin/thick lines represent 50\%/90\% posterior intervals and the squares are the posterior means.}
\label{fig:log_mortality}
\centering
\medskip

\begin{tabular}{cccc}
\includegraphics[width=0.3\textwidth]{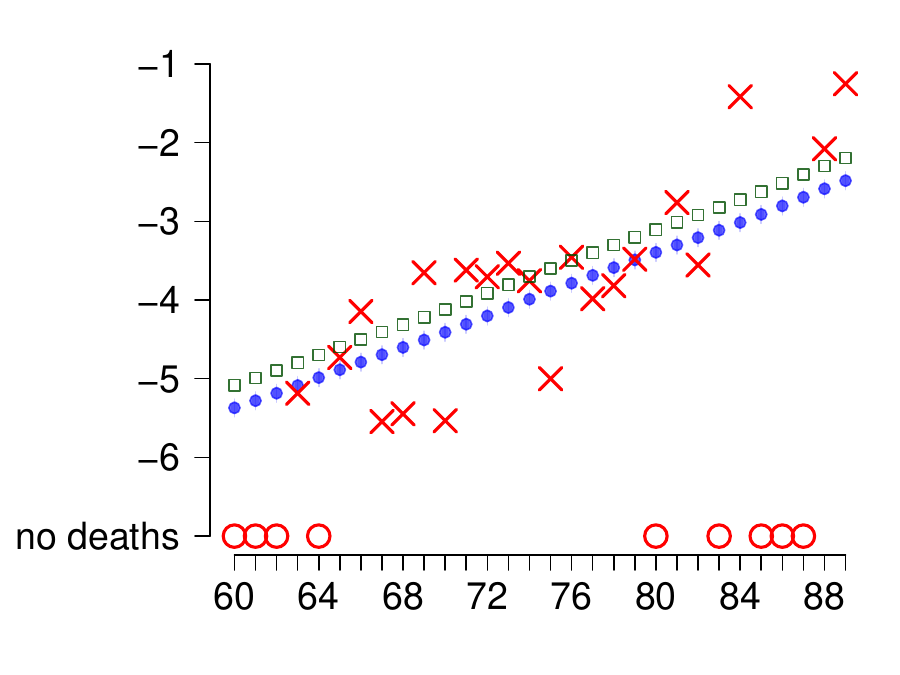} &
\includegraphics[width=0.3\textwidth]{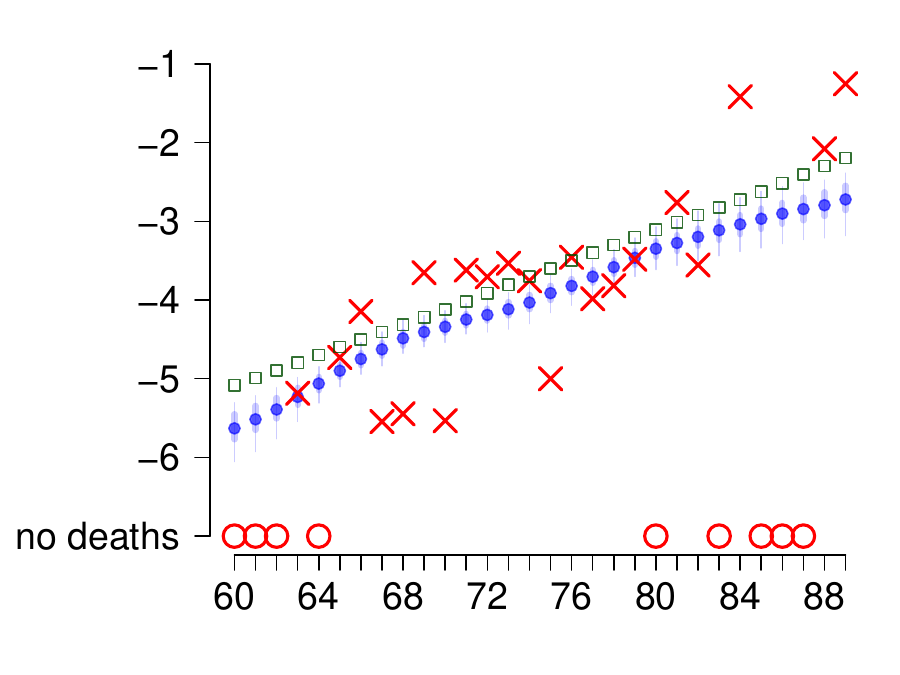} &
\includegraphics[width=0.3\textwidth]{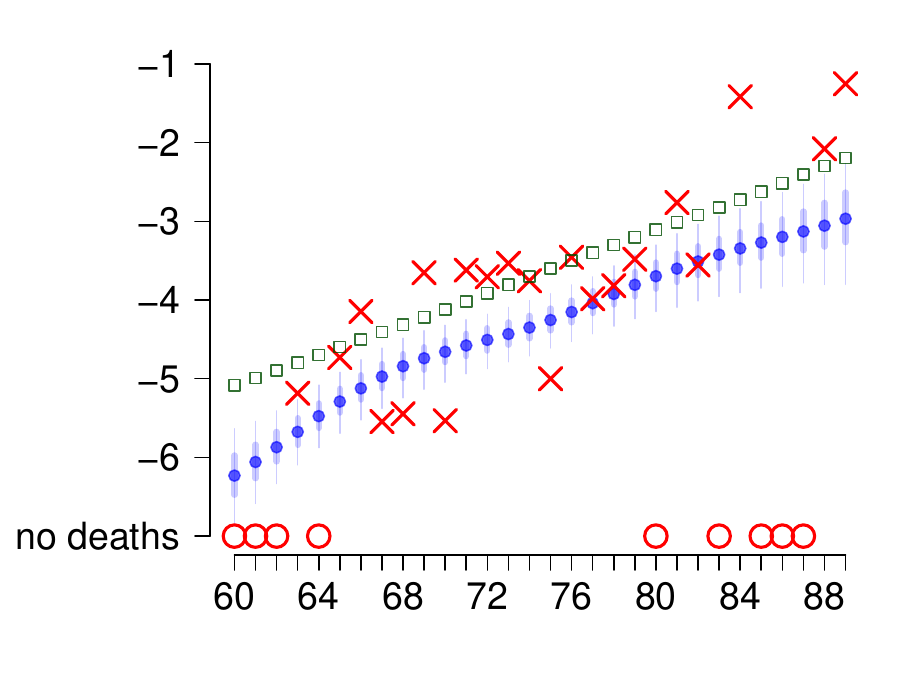}
 \\
 \includegraphics[width=0.3\textwidth]{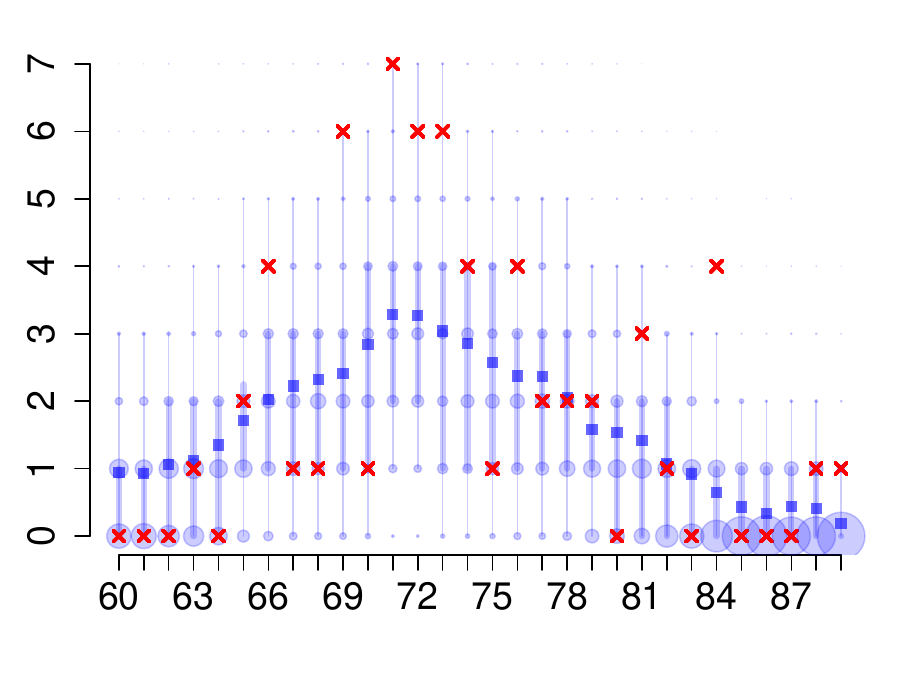} &
\includegraphics[width=0.3\textwidth]{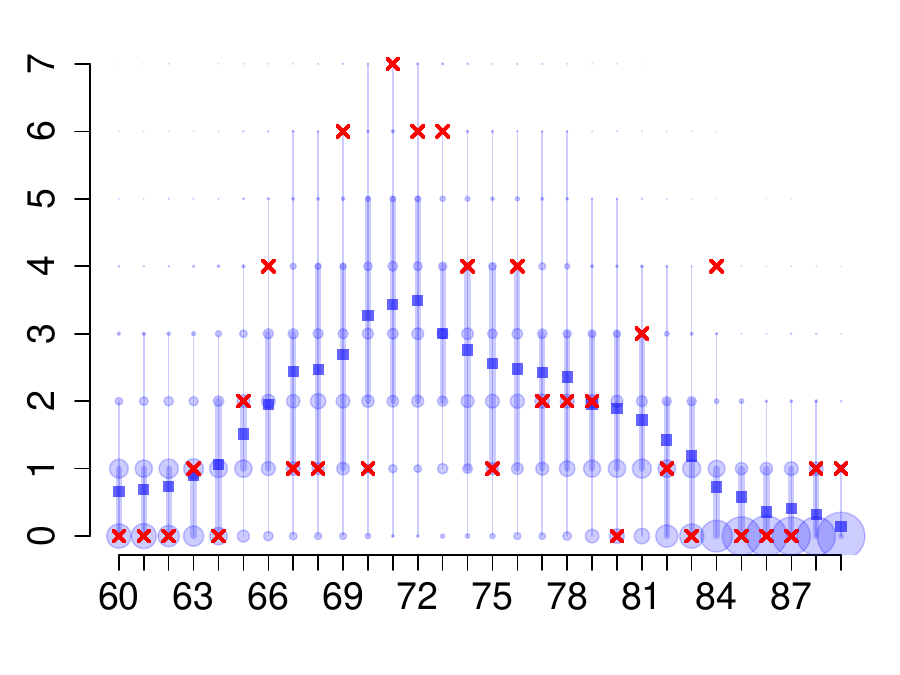} &
\includegraphics[width=0.3\textwidth]{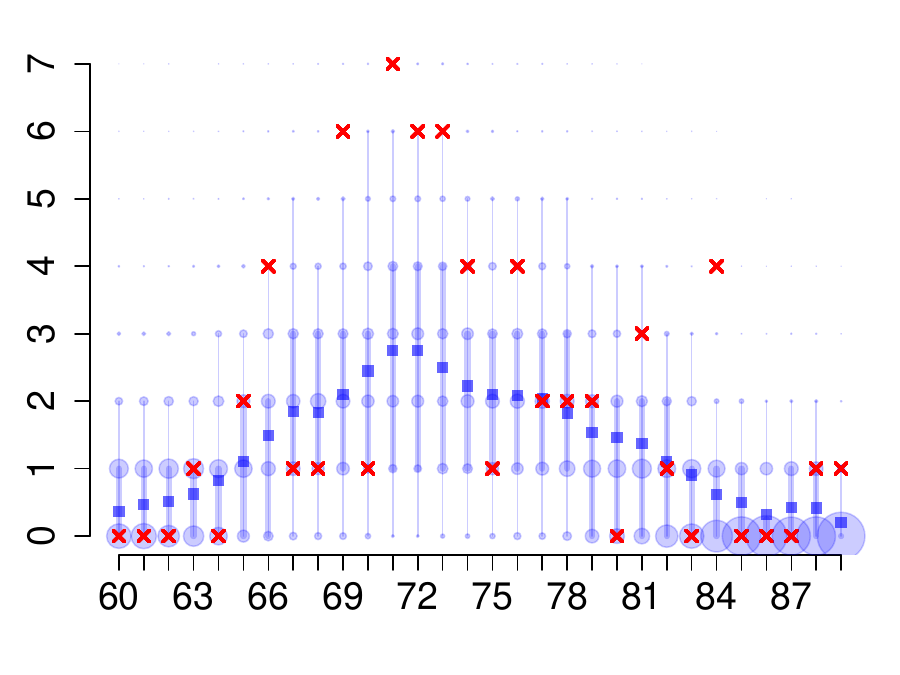} \\
FD-1 & AD-GP & GP-S2 \\
\end{tabular}

\end{figure}

The top row of Figure \ref{fig:log_mortality} shows predicted log-rate (in blue) alongside two other quantities: the country-wide log-mortality (green squares) and the fund's raw mortality $\mu^{raw}_{x,t} =\log( d_{x,t}/E_{x,t})$(red crosses) for the out-of-sample calendar year 2019 and three representative models: FD-1, AD-GP and GP-S2. The bottom row of the Figure displays the respective predictive distribution of age counts. For each age $x$, we show the observed number of deaths $d_{x,t}$ in 2019 as red crosses, and a summary of the posterior predictive distribution in blue. The size of the circles represents the predictive probability assigned to a specific pair (age, number of deaths). The figure also includes the 50\% and the 90\% predictive intervals, represented, respectively, as thin and thick vertical lines. Since death counts are integers, the predictive distribution is discrete, taking on only 2-4 likely values due to the small size of the pension fund.  Figure \ref{fig:log_mortality_appendix} in the Appendix presents the complete results for all 8 models tested. 

Since FD-1 assumes a single deflator for all ages, its resulting log-mortality curve is a parallel shift of the reference mortality. Because the latter is based on the Gompertz law, the resulting log-mortality is linear in $x$. In contrast, the age-dependence of AD-GP and GP-S2 allows for much more flexibility. The raw mortality rates reinforce the pattern of smaller deflators (i.e., larger gaps to national mortality) at ages below 70. Even though the pointwise forecasts (posterior means) present a decreasing pattern in the AD-GP model around ages 85-90, this can not be deemed statistically significant, as the Bayesian interval even for age 90 contains the point estimate for age 85.

While the log mortality rates provide some interpretation on the process driving the pension population's deaths, ultimately the only observable quantity is the actual number of deaths. Overall, the qualitative behavior of the models is largely consistent: for older (above 85) and younger (under 65) ages, all models assign high probability to zero or one deaths; whilst the highest expected number of deaths is at either 71 or 72 years. The only age whose number of deaths is consistently (i.e., across all models) under-estimated is 84, where 4 deaths were observed and all models' maximum a posteriori (MAP) are at zero and the means are lower than one. Other than that, all observations fall at least inside the 90\% predictive intervals and all the models generate a ``triangular'' structure across ages, consistent with the total number of deaths in Figure \ref{fig:funds_exp_pyr}. Note that while we aim to have the predictive intervals for $d_{x,t}$ contain the out-of-sample observations in the bottom row of Figure \ref{fig:log_mortality}, there is no such interpretation for the log-mortality rates in the top row, where the presented uncertainty is about the \emph{latent} mortality curve and not about raw observations (in other words, those error bars are omitting the extra uncertainty of the Negative-Binomial). 

Among the age-dependent models, the log-rates of AD-GP are the smoothest, while AD-FE presents the roughest, consistent with Figure \ref{fig:posterior_theta}. Once again we see the benefit of GP-based models, as one hopes for smooth log-rates. 
Since TD-AR and TD-GP have no age-dependency, their predicted log-mortality curves for 2019 are also parallel shifts of the BRA curves.  As seen in Figure \ref{fig:posterior_theta}, the predicted point estimate for the 2019 deflators is very similar for TD-AR and TD-GP, making the resulting log-mortality curves almost indistinguishable in Figure \ref{fig:log_mortality_appendix}.
Predicted log-mortality rates for the single population models (GP-S1 and GP-S2) are found in the bottom row of Figure \ref{fig:log_mortality_appendix}. Similar to the age-dependent GP model, GP-S1 produces mortality rates that are decreasing between ages 80 to 90. In the GP-S2 model, which has both time and age dependence, this issue is not observed and the predicted log-rates appear to be more plausible. Of note, the rates inferred by GP-S2 look like a smoothed version of those found by AD-AR, where the fund's rates are much lower than the national ones for younger ages (60 to 70) but show signs of convergence for older ages (80-90).

\subsection{Specific model parameters}
All model hyperparameters proposed are easily interpretable; Figure \ref{fig:hiperpost_appendix} in the Appendix  shows the posteriors for AD-AR, AD-GP, TD-AR and TD-GP models. Figure \ref{fig:parameters_gps1_gps2} presents  the posterior distributions of key model parameters for the GP-based single population mortality models GP-S1 and GP-S2. Qualitatively comparing the posteriors (blue densities) with the assigned priors (red densities) we confirm that our priors are, indeed, weakly informative, as claimed in the previous section.  

\textbf{Presence of Age-correlation/persistence:} In the two autoregressive models AD-AR and TD-AR, 
a prior distribution is only directly assigned to the slope $\rho^i$ whereas $\mu^i$'s are induced by the former. As seen in AD-AR's results, the posterior distributions for these two parameters end up being essentially equivalent (apart from a change in scale). The posterior mean of $\rho^i$ for AD-AR is $0.78$ with a 90\% posterior quantile interval of $[0.28; 0.99]$.
The slope parameter is related to deflator persistence. Recall that the random walk (RW) model is a special case of the autoregressive (AR) model, in which the slope parameter $\rho$ is equal to $1$.  Thus, an AR model exhibits higher persistence when its slope parameter is closer to $1$, but the process reverts to its mean fairly quickly. We observe that the $\rho$ parameter of TD-AR is a bit lower than that of AD-AR,  meaning that the age-dependent deflators  have ``longer memory'' than time-dependent deflators.

\begin{figure}[!ht]
\caption{Posterior distribution of GP lengthscales: $\phi_{ag}^i$ in AD-GP, GP-S1 and GP-S2 (\emph{top row}) and $\phi_{yr}^i$ in TD-GP, GP-S2 (\emph{bottom row}) models. The red curve indicates the prior density as listed in Table \ref{tab:modSOGP}, the vertical solid (dashed) lines denote the posterior mean (90\% posterior interval).  \label{fig:lengthscale-posteriors} }
\hspace*{-6pt}\begin{tabular}{cccc}
$\phi_{ag} \negmedspace\negmedspace$ & \raisebox{-.5\height}{\includegraphics[width=0.29\textwidth, trim=0.2in 0in 0.2in 0in]{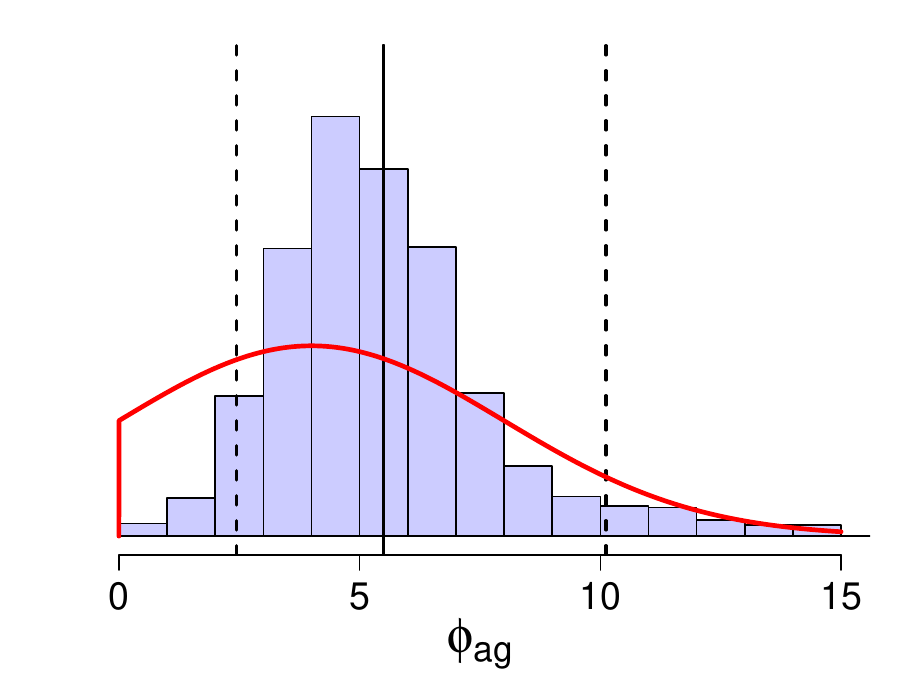}} & \raisebox{-.5\height}{\includegraphics[width=0.29\textwidth, trim=0.2in 0in 0.2in 0in]{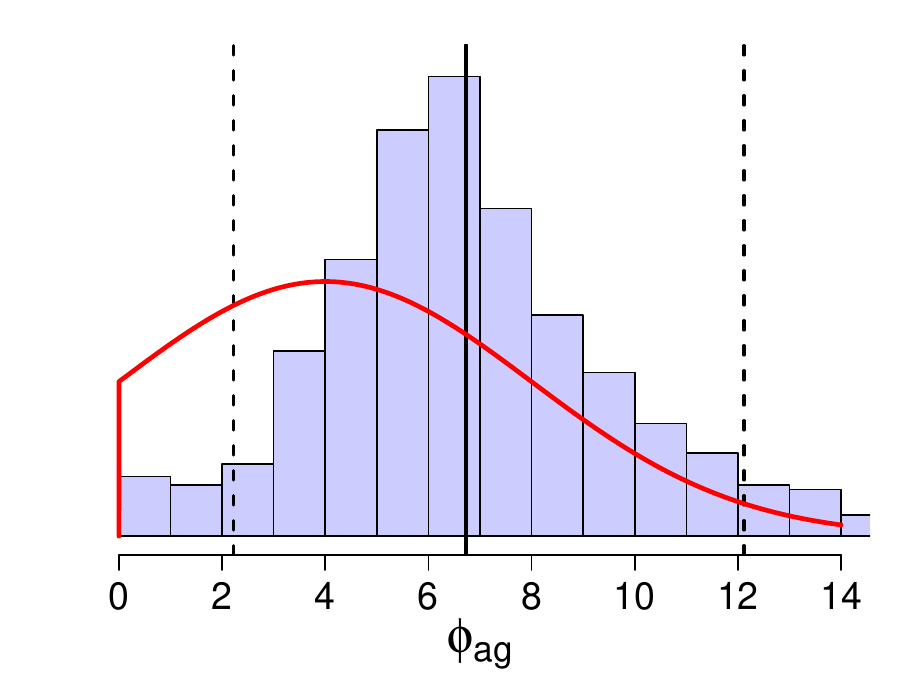}} & 
\raisebox{-.5\height}{\includegraphics[width=0.29\textwidth, trim=0.2in 0in 0.2in 0in]{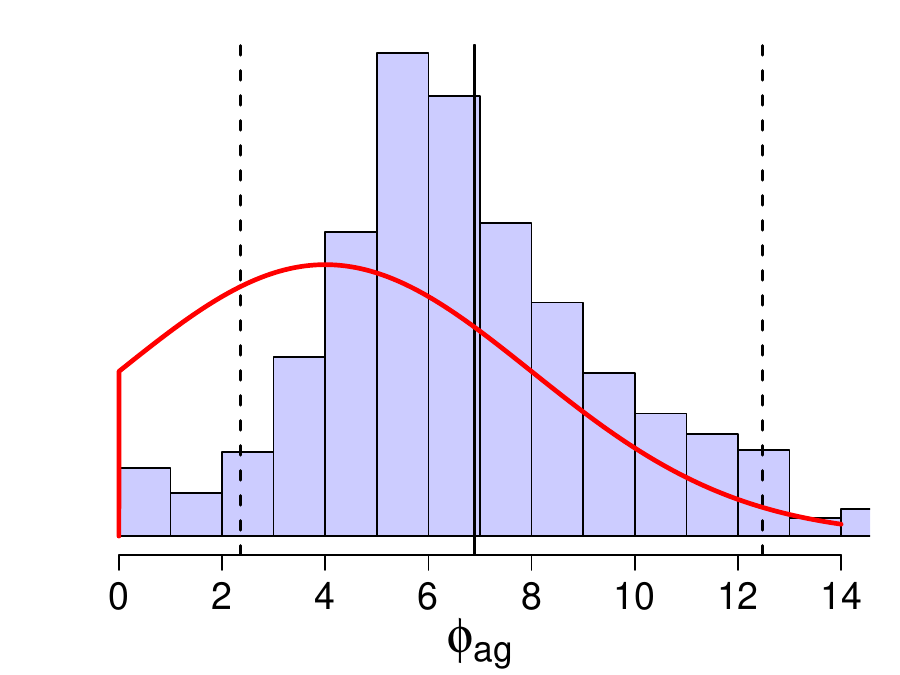}}   \\
 & AD-GP & GP-S1 & GP-S2 \\
 $\phi_{yr}\negmedspace\negmedspace$ & \raisebox{-.5\height}{\includegraphics[width=0.29\textwidth, trim=0.2in 0in 0.2in 0in]{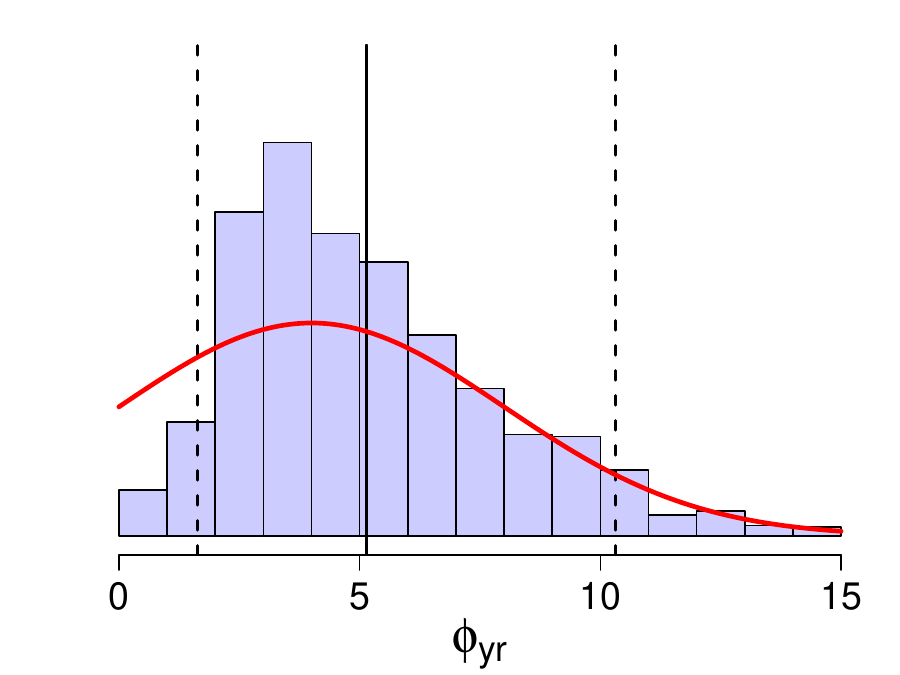}} &  & 
 \raisebox{-.5\height}{\includegraphics[width=0.29\textwidth, trim=0.2in 0in 0.2in 0in]{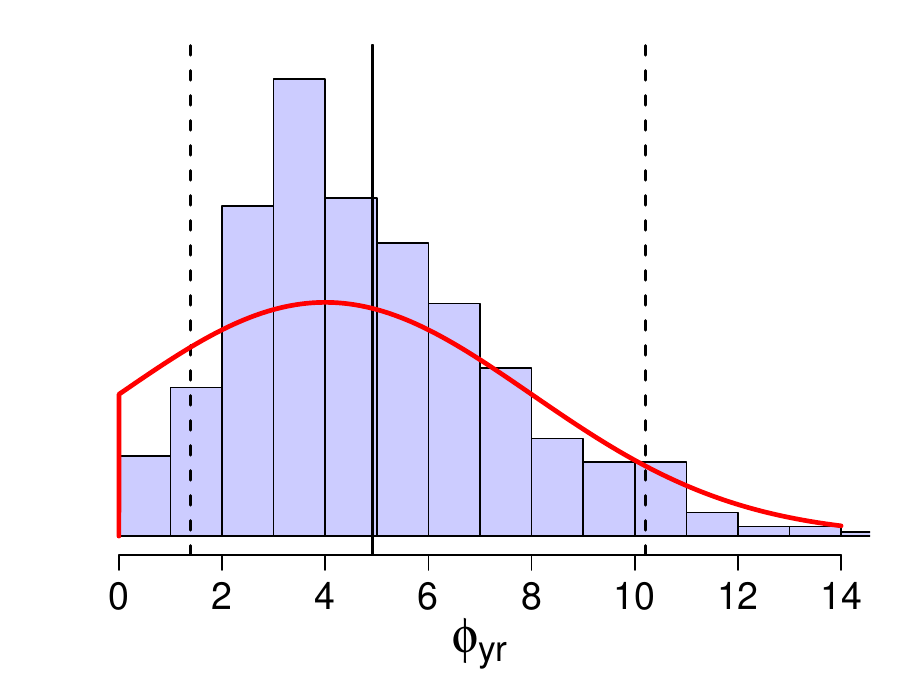}} \\
 & TD-GP &  &  GP-S2 
\end{tabular}
\end{figure}

\textbf{Age and Year Lengthscales:} Figure \ref{fig:lengthscale-posteriors} shows the posteriors of $\phi_{ag}$ and $\phi_{yr}$ across the various GP-based models.
For the age lengthscale, the estimated 90\% posterior ranges are $[2.42; 10.09]$ (posterior mean $5.52$) for AD-GP, $[2.22; 12.11]$ (mean  $6.73$) for GP-S1 and  $[2.36; 12.47]$ (mean 6.90) for GP-S2. This can be interpreted as fusing information across about $2\phi_{ag} \simeq 10$-age groups when estimating a given deflator. This estimate is consistent across the 3 models and matches other age lengthscales in the literature, which globally reflect the idea of a generation being about 15 years.  We note that the posterior is very wide, so the models are not able to definitively infer the Age-persistence of deflators in the pension fund. 
While the dependence structure of a GP with squared-exponential kernel is quite different from the autoregressive one, as a point of comparison, the GP structure implies that $cor(\theta_x, \theta_{x+1}) =  \exp(-1/(2 \phi^2_{ag}))$ while the AR gives $cor(\theta_x, \theta_{x+1}) = \rho$. Thus, $\rho=0.9$ roughly matches $\phi_{ag}=2.2$ and $\rho=0.99$ translates to $\phi_{ag}=7.1$.

The estimated temporal lengthscale $\phi_{yr}^i$ has posterior mean 5.1 in TD-GP (posterior 90\% quantile interval of $[1.64; 10.30]$) and 4.90 in GP-S2, again showcasing consistency. Furthermore, note that
relative to the age-dependent deflators, $\phi_{yr} < \phi_{ag}$, i.e., the temporal correlation is weaker than across ages, which is intuitive and consistent with prior research. 

\textbf{Parametric trends:} the estimated age trend is $\beta_{ag} \simeq 0.12$ for GP-S1 and $\beta_{ag} \simeq 0.11$ in GP-S2. The estimated year trend in GP-S2 is $\beta_{yr} \simeq -0.078$ ($[-0.15; -0.01]$), which is large but unavoidable.
This means that in 5 (9) years, the mortality rates will be approximately 30\%(50\%) lower, a rather unrealistic longevity improvement. The same coefficient estimated for the Brazilian national population is $-0.016$ considering same ages and calendar years. This relative gap in mortality improvement cannot persist for long periods of time but could well hold over a decade or so, implying that the GP-based models are not suitable for long extrapolation in Year. 
Mortality rates curves obtained with time deflators are increasing and smoother than the ones for the age deflators models, since the cross section age structure of the reference population is kept for the pension fund population.

By construction, the estimates of $\beta_0$ should be consistent across models, corresponding to the estimated mortality at age 59. As shown in the Figures, we get $\beta_0 \in [-7,-4]$ in all cases with very similar posterior means. One may notice a difference in posterior means of $\beta_0$ across the two GP models, with GP-S1 showing a lower $\beta_0$ compared to the GP-S2. This is due to the use of a year covariate in GP-S2. Since GP-S1 model does not take into account a linear decrease on log-mortality rates on time, $\beta_{0}$ is lower in this model to best fit lower mortality rates as time goes by.

When directly modeling mortality rates, one may get results for smaller sub populations exhibiting unrealistic trends over the medium to long term. Within our sample, a notably high annual improvement rate was observed (see the posterior histograms of $\beta_{yr}$ in Figure \ref{fig:parameters_gps1_gps2}). Consequently, it appears advantageous to forecast mortality within (small) pension funds by employing a reference population that has been adjusted by deflators. Utilizing age deflators allows for some adaptability in addressing variations across different age groups in mortality patterns between the pension fund and the reference population, while still reflecting the improvements over time of the latter. Conversely, time deflators accommodate differences in temporal improvements between the two populations but offer limited flexibility in addressing idiosyncratic variations across different age groups. In this context, one may suggest to employ deflators on both coordinates: age and time. However, employing such an approach may offer minimal or no advantage compared to directly modeling the mortality rates of the pension fund.

\subsection{Out-of-sample performance}

\begin{figure}[ht!]
\caption{Predicted pension fund 2019 log-mortality rates for Ages 60-89 based on the proposed GP-based models trained on 2013--2018. x's indicate the observed raw mortality rates in 2019; note that for some ages there were zero recorded deaths, shown as circles at the bottom of the plot.}
\centering
\includegraphics[width=0.8\textwidth]{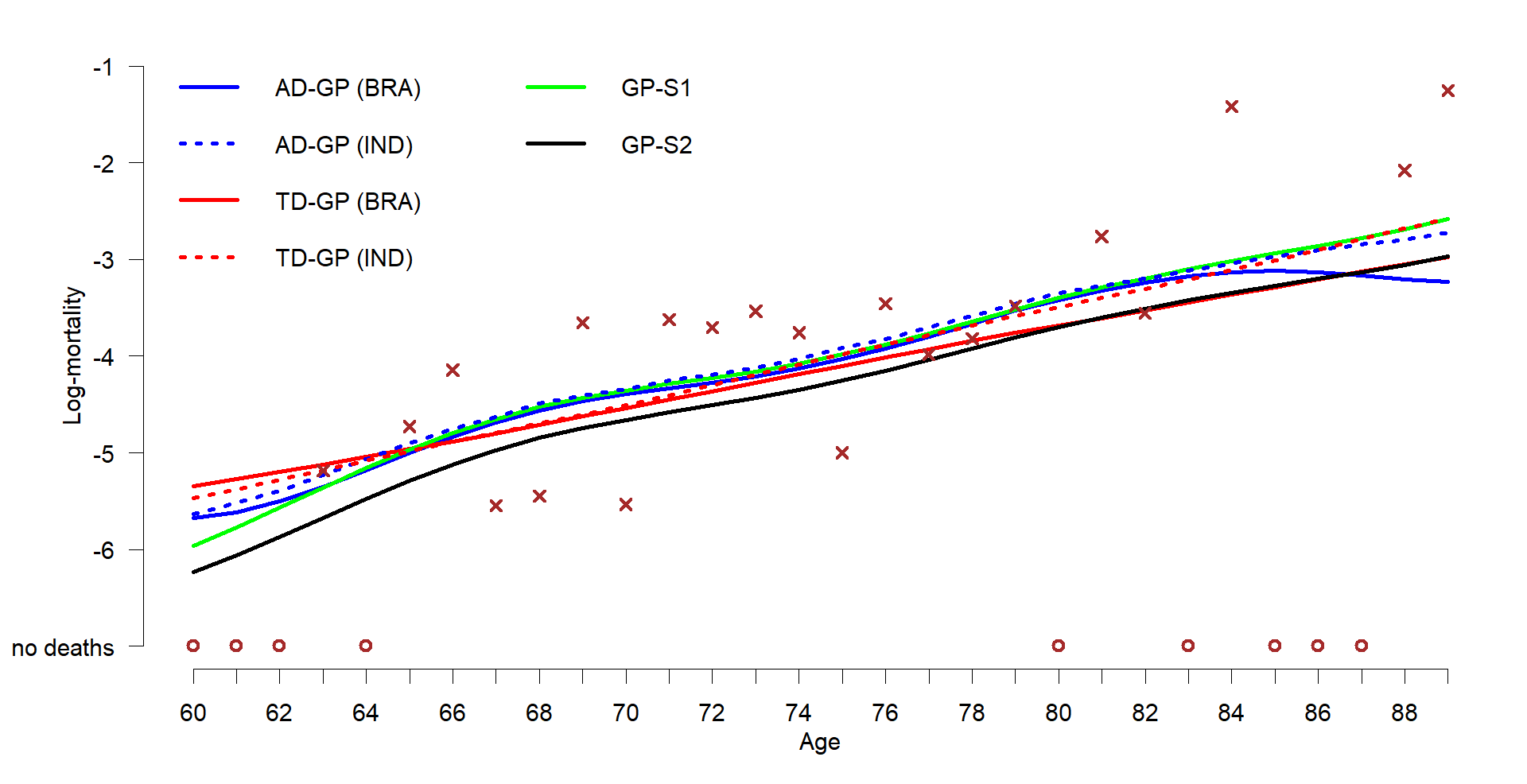}
\label{fig:Fitted_mort_curves}
\end{figure}

Figure \ref{fig:Fitted_mort_curves} presents predicted 2019 pension fund mortality across the GP-based models. We show out-of-sample projects of the  AD-GP and TD-GP models that use the national mortality table as a reference ($i=BRA$), as well as the forecasted  log-mortalities for GP-S1 and GP-S2. Comparing models AD-GP and TD-GP to GP-S1/2 can shed light on the value of having a reference population and borrowing information from it, compared to a direct approach for the pension data.

The reader is reminded that, under the AD-GP, the calendar year dynamics of the pension fund population is determined by the calendar year dynamics of reference population. On the other hand, for time-dependent deflator models, the age structure of the pension fund population is based on (deflated) reference populations and there is effectively a single deflator across ages for a given year, but the time dynamics of the pension fund population is more flexible. 

Under the TD-GP model, when forecasting the pension fund mortality, it is necessary to first forecast the reference population mortality rates. For the sample period, we used the available and published mortality rates, but this task may be done for non-observed future years by an adequate model for national populations such as, for example, the  \cite{lee1992modeling} model.

We observe that the 2019 projections are quite similar across the four models in Figure~\ref{fig:Fitted_mort_curves}, with differences largely confined for the edges (ages less than 65 or over 85). GP-S1 and AD-GP almost exactly match each other in the middle of the range, and give the best fit relative to the realized mortality. We also note that GP-S1 and GP-S2 are effectively parallel offsets of each other, i.e., yield the same projected age structure up to a constant that can be attributed to the year-trend embedded in GP-S2.

\section{Model Assessment}\label{sec:discussion}

In order to more rigorously compare the models, we use probabilistic performance metrics based on evaluating the model-based predictive distributions relative to the observed mortality data.
We refer to \cite{czado2009predictive} for the evaluation of probabilistic forecasts  for count data. We consider two proper scoring rules that are applicable
for both Bayesian and classical, as well as parametric or non-parametric settings with discrete outcomes. Given a predictive probability mass function $p_{x,t}(\cdot)$, and the respective observed death count $d_{x,t}$ we first compute the log-score
\begin{eqnarray}\label{eq:log-score}
\mathrm{score}_{\log} = - \frac{1}{N}\sum_{x,t} \log p_{x,t}(d_{x,t})
\end{eqnarray}
\noindent where $N$ is the number of observations. 
Since our predictive distributions $p_{x,t}(\cdot)$ are only available through simulation, so are the logarithmic scores. The best fit should present the lowest $\mathrm{score}_{\log}$.

The other measure applied to assess the performance of the tested models is the rank probability score (RPS):
\begin{eqnarray}\label{eq:rps}
\mathrm{score}_{RPS} = \frac{1}{N} \sum_{x,t} \bigg[ \sum_{k=1}^{\bar{d}} \big(P_{x,t}(k) - \mathbb{I}(d_{x,t} \le k)\big)^2 \bigg].
\end{eqnarray}

\noindent Above $P_{x,t}(k) = \mathbb{P}(d_{x,t} \le k) = \sum_{\ell=0}^k p_{x,t}(\ell)$ is the predictive CDF of $d_{x,t}$ for age $x$ and year $t$. Empirically, we restrict to $k \le \bar{d} = 10$ given the single-digit distribution of observed age-specific deaths in our pension fund, see Figure~\ref{fig:log_mortality} bottom row. Thus,  $\mathbb{P}(d_{x,t} \le 10) \simeq 1$ for all $x,t$. This predictive CDF is compared to the realized observation $d_{x,t}$; lower $\mathrm{score}_{RPS}$ again indicates better fit.

We also computed mean predictive absolute error (MAE) metric, though given that $d_{x,t}$ has a highly skewed asymmetric distribution, MAE is not very suitable and primarily focuses on Ages 70--80 where predictive variance of $d_{x,t}$ is highest. Most models end up with very similar MAEs of about 1.01, except for FD-0 which has a MAE of 1.61 (with BRA reference) and 1.11 (with IND reference population). 

Table \ref{tab:modelsBinNeg} presents the results for each one of the models. The out- and in-sample results in the table are the mean of the yearly log and RPS scores for a leave-one-out cross-validation in the sample period 2013-2019 and ages 60-89. In each year, $N=30$ in equations \ref{eq:log-score} and \ref{eq:rps} for the out-of-sample and $N=210-30=180$ for  the in-sample metrics.
Note that because FD-0 does not provide a probabilistic forecast, it is omitted from Table~\ref{tab:modelsBinNeg}. The results confirm that FD-1 is vastly outperformed by all other models, and that only capturing the year-trend (TD-AR and TD-GP) is similarly inadequate. The best scores are achieved by AD-AR and AD-GP, as well as the GP-S1 model, which can be declared the overall winner. GP-S2 appears to strongly overfit, yielding the best in-sample performance but poor out-of-sample scores.

\begin{table}[!ht]
\centering
\begin{tabular}{ c|c|c|c|c  }
\hline   

\multirow{2}{*}{Model} & \multicolumn{2}{c|}{RPS}  & \multicolumn{2}{c}{Log-score} \\ \cline{2-5}
 & Out-of-sample & In-sample & Out-of-sample & In-sample \\ \hline

FD-1   & 0.662 & 0.770 & 1.428 & 1.593 \\

AD-FE  & 0.665 & 0.730 & 1.433 & 1.543 \\

AD-AR  & 0.654 & 0.703 & 1.422 & 1.506 \\

AD-GP  & 0.654 & 0.717 & 1.425 & 1.559 \\

TD-AR  & 0.674 & 0.772 & 1.452 & 1.583 \\

TD-GP  & 0.662 & 0.768 & 1.440 & 1.590 \\

GP-S1  & {\bf 0.649} & 0.718 & {\bf 1.413} & 1.548 \\
GP-S2  & 0.667 & {\bf 0.620} & 1.428 & {\bf 1.377} \\ \hline 
\end{tabular}
\caption{\label{tab:modelsBinNeg} Mean of the yearly performance indexes \eqref{eq:log-score}-\eqref{eq:rps} for leave-one-out cross-validation across years 2013-2019, ages 60-89 for Males. $BRA$ as reference population, except for models GP-S1 and GP-S2. Out-of-sample and in-sample results considering models described in Table \ref{tab:modSOGP} and Negative Binomial likelihood. Bolded numbers indicate the best-performing model for each metric.}
\end{table}

There is a regulatory requirement in Brazil for pension funds to perform a consistency test on the mortality table used for calculating technical provisions. This consistency test is a chi-square goodness-of-fit test that compares the observed number of deaths for a given set of ages with the expected number of deaths as defined by a mortality table. Accordingly, we conducted these tests using the AT-2000M table, as well as the mortality rates estimated by our proposed AD-GP and GP-S1 models. The resulting p-values were: $25.67\%$ for the AT-2000M table, $60.37\%$ for the AD-GP model, and $59.80\%$ for the GP-S1 model.

Although none of the tables or models led to rejection of the null hypothesis—that is, all are statistically consistent with the observed data—the proposed models demonstrated a notably better fit. The higher p-values associated with the AD-GP and GP-S1 models suggest that they more accurately reflect the underlying mortality patterns of the sample, thereby reinforcing their potential applicability in actuarial practice.

\section{Conclusion}
Pension fund liability management requires the use of consistent, realistic and up-to-date mortality rates for the fund, as well as their future projection for cashflow discounting. Practicing actuaries tend to rely on national population data for these purposes, in particular to use external benchmarks by official statistics offices that regularly update national table projections. To capture the basis risk between fund mortality and national data we explored a variety of deflator-based models. Modeling the sub-population as a deflated version of the national table leverages the stability and credibility of the national projections, while still capturing the specific characteristics of the sub-population. 

This work demonstrates that Gaussian Process (GP) based models add a flexible and robust alternative to the menu of sub-population mortality models available for actuarial applications. In the examples tested, GP-based models generated smooth mortality curves and were able to capture the variability in age deflators, outperforming more rigid parametric approaches. The ability of GPs to borrow strength across ages and years is particularly valuable in small populations, where data is sparse and traditional models may overfit or underfit.

The analyzed dataset is novel in the actuarial literature and highlights the high inequality in mortality rates across different groups in Brazil, where national mortality rates can be almost double those of a selected sub-population, such as participants of a pension fund. The sub-population modeling approach, supported by GP-based deflators, proved effective in borrowing information from the reference population and producing credible mortality nowcasts even in the presence of significant heterogeneity and single-digit observed death counts. Indeed, the motivation for this study came from practitioners who questioned the adequacy of standard annuity tables, such as the AT-2000, for their specific populations. Our approach, based on the actual experience of Brazilian pension funds and validated through consistency tests required by the regulator, provides a better fit and a more transparent methodology for practitioners and regulators alike.

Another feature we encountered was that in the case study discussed, the observed yearly mortality improvements were unusually high, reflecting particularities of the population under analysis. In such situations, we recommend the use of informative priors, naturally afforded by our framework, that borrow information from the projections of the reference population, thus avoiding unrealistic extrapolations and ensuring more stable forecasts.

\bibliographystyle{apalike}
\bibliography{subpops}

\newpage

\appendix

\section{Models}

\begin{table}[ht!]
\centering
\begin{tabular}{ c|c|c|c  }
Model & Likelihood & Predictor & Prior distributions \\ 
  & *where $i = \{BRA, IND\}$ and  &  &  \\ \hline
  
FD-0& $d_{x,t} = m^{i}_{x,t} E_{x,t}$ & --- & --- \\
    &  &   &  \\ \hline

FD-1& $ d_{x,t} \sim \NB(e^{\theta^i} \cdot m^{i}_{x,t} E_{x,t}, \omega^i)$ & --- & $\theta^i \sim \cN(-0.5,0.5^2)$\\

 &  &  & $\omega^i \sim \cN(0,1^2)I_{(0,\infty)}$ \\ \hline

\multicolumn{4}{c}{Age-dependent deflators (AD)} \\ \hline

AD-FE & $d_{x,t} \sim \NB(e^{\theta^i_{x}} \cdot m^{i}_{x,t} E_{x,t}, \omega^i)$ & --- & $\theta^i_{x} \sim \cN(-0.5,0.5^2)$\\

 &  &  & $\omega^i \sim \cN(0,1^2)I_{(0,\infty)} $ \\ \hline

AD-AR & $  d_{x,t} \sim \NB(e^{\theta^i_{x}} \cdot m^{i}_{x,t} E_{x,t},\omega^i)  $ & $\theta^i_{x} = \mu^i + \rho^i \theta^i_{x-1} + \epsilon^i_{x}$ & $\rho^i \sim \cN(1,1^2)I_{(0,1)}$ \\

&   & $\epsilon^i_{x} \sim \cN(0,0.5^2(1-\rho^i)^2)$ & $\theta^i_{60} \sim \cN(-0.5,0.5^2)$ \\

&   & $\mu^i = - 0.5 \times (1 - \rho^i)$ & $\omega^i \sim \cN(0,1^2)I_{(0,\infty)}$ \\  \hline

AD-GP & $ d_{x,t} \sim \NB(e^{\theta^i_{x}} \cdot m^{i}_{x,t} E_{x,t},\omega^i) $ & $ \theta^i_{x} \sim GP(-0.5, c^i_{ag}(\cdot,\cdot)) $ & $ (\sigma^i)^2 \sim \cN(0.5,0.5^2)I_{(0,\infty)} $ \\

 &  &  &  $ \phi^i_{ag} \sim \cN(4,4^2)I_{(0,\infty)} $ \\

 &  &  & $\omega^i \sim \cN(0,1^2)I_{(0,\infty)} $ \\ \hline

\multicolumn{4}{c}{Time-dependent deflators (TD)} \\ \hline

TD-AR & $d_{x,t} \sim \NB(e^{\theta^i_{t}} \cdot m^{i}_{x,t} E_{x,t},\omega^i)$ & $\theta^i_{t} = \mu^i + \rho^i \theta^i_{t-1} + \epsilon^i_{t}$ & $\rho^i \sim \cN(1,1^2)I_{(0,1)}$ \\

 &   & $\epsilon^i_{t} \sim \cN(0,0.5^2(1-(\rho^i)^2)$ & $\omega^i \sim \cN(0,1^2)I_{(0,\infty)}$ \\
 &   & $\mu^i = -0.5 \times (1-\rho^i)$ & $\theta^i_{2013} \sim \cN(-0.5,0.5^2)$ \\ \hline

TD-GP  & $ d_{x,t} \sim \NB(e^{\theta^i_{t}} \cdot m^{i}_{x,t} E_{x,t},\omega^i) $ & $ \theta^i_{t} \sim GP(-0.5, c^i_{yr}(\cdot,\cdot)) $ & $(\sigma^i)^2 \sim \cN(0.5,0.5^2)I_{(0,\infty)} $ \\

 &  &  &  $ \phi^i_{yr} \sim \cN(4,4^2)I_{(0,\infty)} $ \\

 &  &  & $\omega^i \sim \cN(0,1^2)I_{(0,\infty)} $ \\ \hline

\multicolumn{4} {c} {Direct population mortality modeling} \\ \hline

GP-S1     & $d_{x,t} \sim \NB(e^{\psi_{x}} E_{x,t}, \omega)  $ & $ \psi_{x} \sim GP(\mu_{ag}(x), c_{ag}(\cdot, \cdot))  $
 &  $ \beta_{0} \sim \cN(-5,1^2) $ \\

 & & $ \mu_{x} = \beta_{0} + \beta_{ag} (x-60) $ &  $ \beta_{ag} \sim \cN(0.1,0.1^2) $ \\
 & & & $ \sigma^2 \sim \cN(0.5,0.5^2)I_{(0,\infty)} $ \\
 & & & $ \phi_{ag} \sim \cN(4,4^2)I_{(0,\infty)} $ \\
 & & & $\omega \sim \cN(0,1^2)I_{(0,\infty)} $ \\ \hline

GP-S2     & $ d_{x,t} \sim \NB(e^{\psi_{x,t}} E_{x,t}, \omega) $ & $\psi_{x,t} \sim GP(\mu_{x,t}, c_{ag,yr}(\cdot,\cdot))$ &  $ \beta_{0} \sim \cN(-5,1^2) $ \\

 & & $ \mu_{x,t} = \beta_{0} + $
 &  $ \beta_{ag} \sim \cN(0.1,0.1^2) $ \\

 & & $\beta_{ag} (x-60) + $ & $ \beta_{yr} \sim \cN(0,0.1^2) $ \\

 & & $\beta_{yr} (t-2013)$ & $ \sigma^2 \sim \cN(0.5,0.5^2)I_{(0,\infty)} $ \\

 & & & $ \phi_{ag} \sim \cN(4,4^2)I_{(0,\infty)} $ \\

 & & & $ \phi_{yr} \sim \cN(4,4^2)I_{(0,\infty)} $ \\
 
 & & & $\omega \sim \cN(0,1^2)I_{(0,\infty)} $ \\ \hline

\end{tabular}
\caption{\label{tab:modSOGP} Model specifications for the pension fund number of deaths.}
\end{table}

\newpage

\section{Supplementary Figures} \label{sec:appendix}

\begin{figure}[!ht]
\caption{Predicted pension fund 1's log-mortality rates for 2019 (blue) induced by different models, reference population (BRA) mortality (green squares) and raw mortality (red crosses) at 2019 for the Male population for different ages in the horizontal axis.}
\label{fig:log_mortality_appendix}
\centering
\medskip

\begin{tabular}{cccc}
\includegraphics[width=0.3\textwidth]{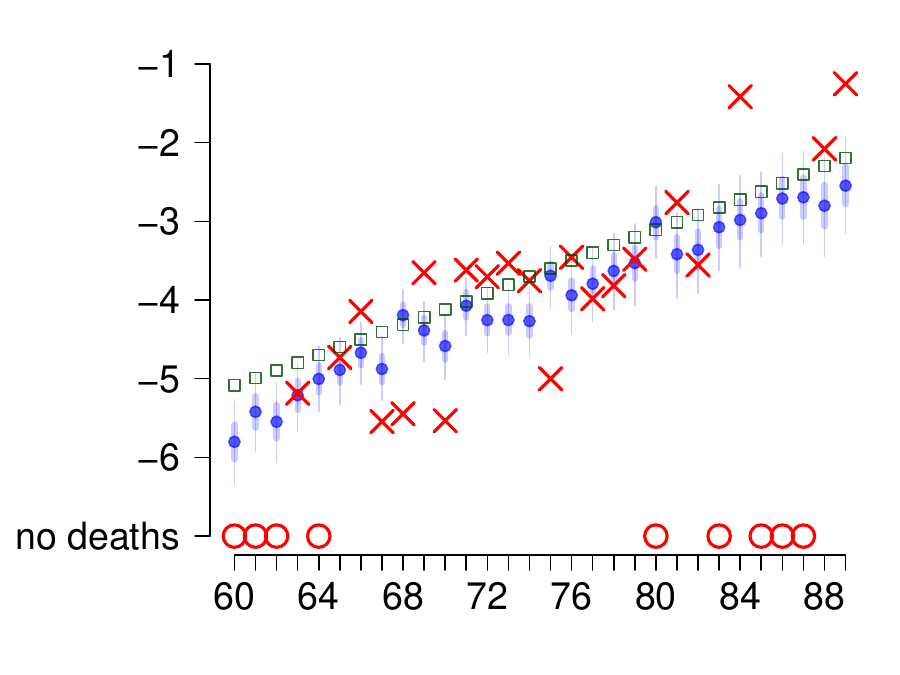} &
\includegraphics[width=0.3\textwidth]{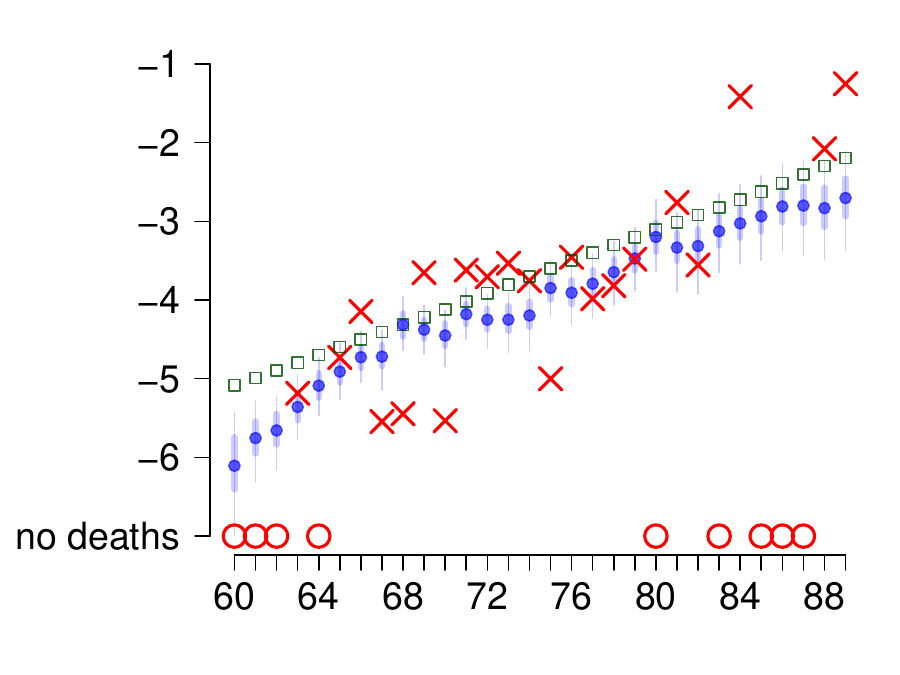} &
\includegraphics[width=0.3\textwidth]{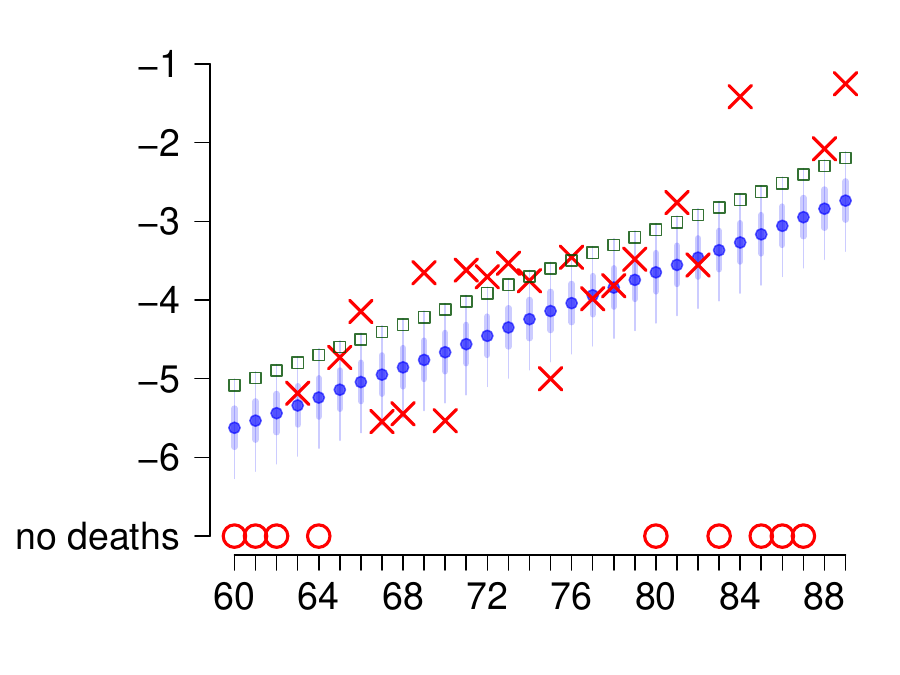} \\
AD-FE & AD-AR & TD-AR \\
\includegraphics[width=0.3\textwidth]{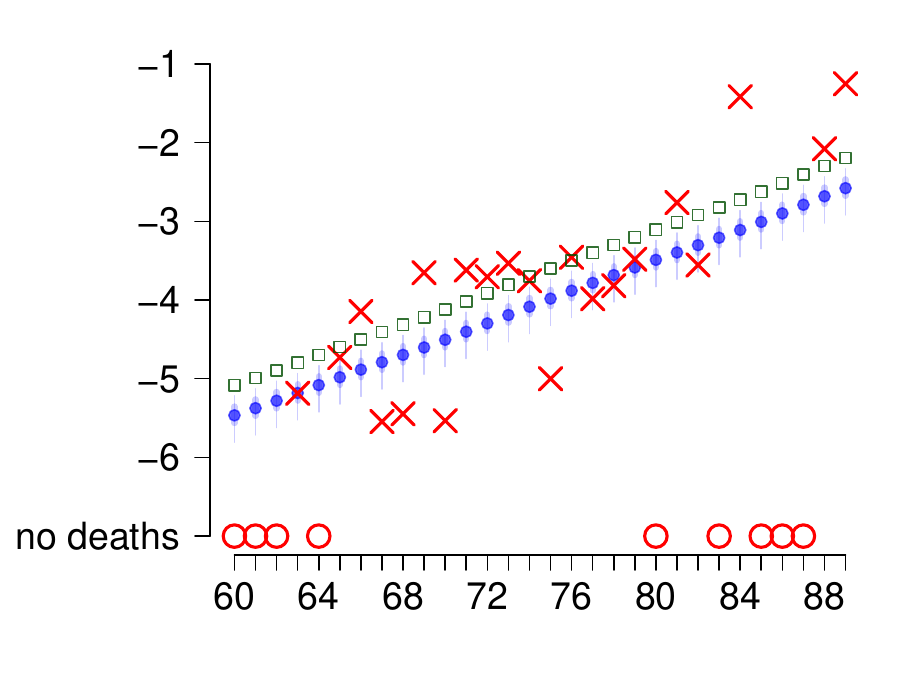} &
\includegraphics[width=0.3\textwidth]{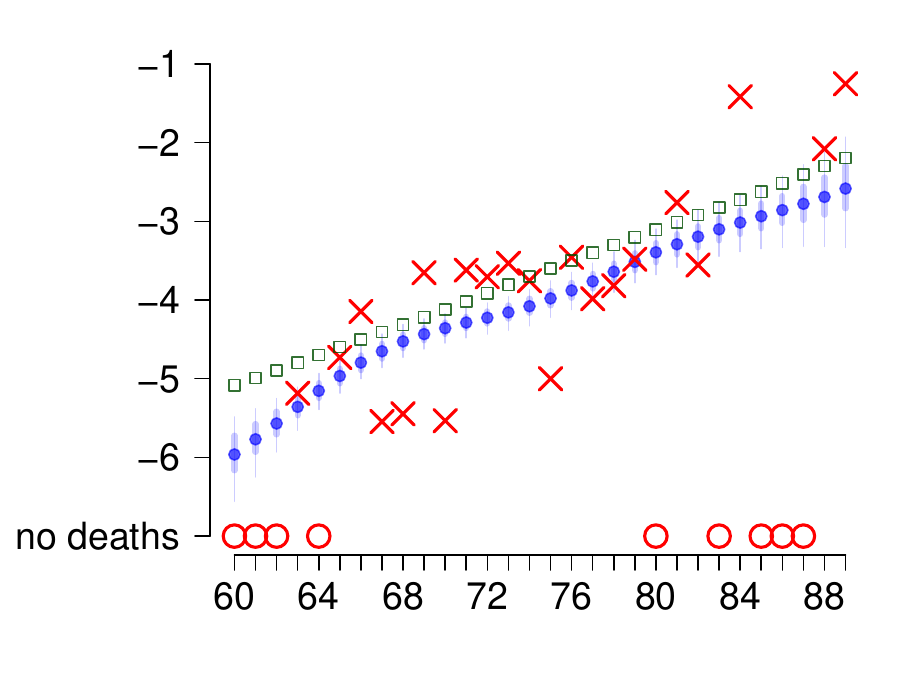} &
 \\
TD-GP & GP-S1 & \\
\end{tabular}

\end{figure}

\begin{figure}[!ht]
\caption{Predicted pension fund 1's number of deaths for 2019 (blue) induced by different models and observed number of deaths (red crosses) for the Male population for different ages in the horizontal axis. Circle sizes represent the probability assigned to the corresponding number of deaths, thin/thick lines represent 50\%/90\% posterior intervals and the squares are the posterior means.}
\label{fig:prediction_deaths_appendix}
\centering
\medskip

\begin{tabular}{ccc}
\includegraphics[width=0.3\textwidth]{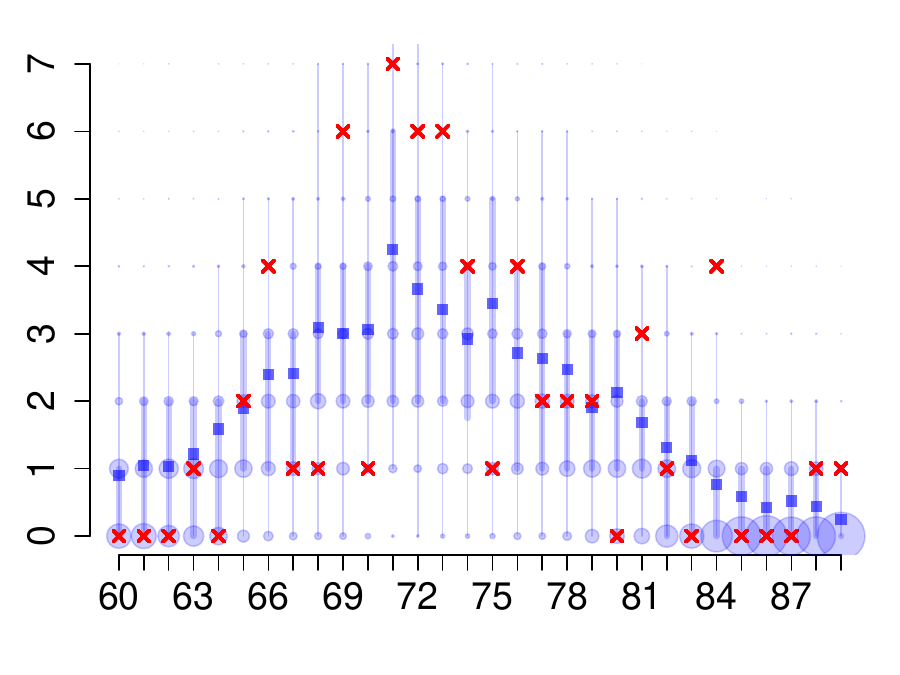} &
\includegraphics[width=0.3\textwidth]{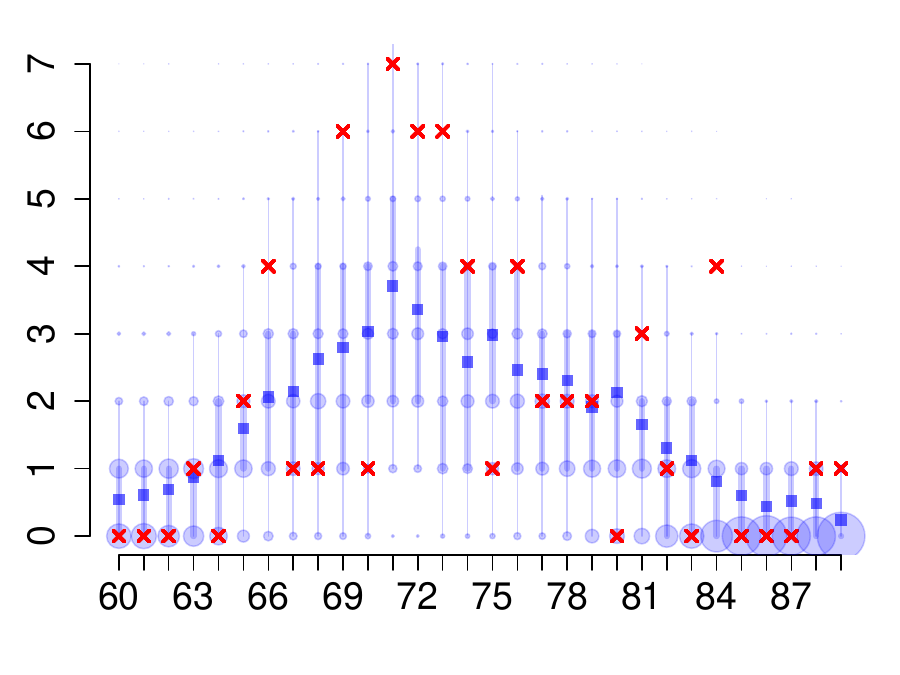} &
\includegraphics[width=0.3\textwidth]{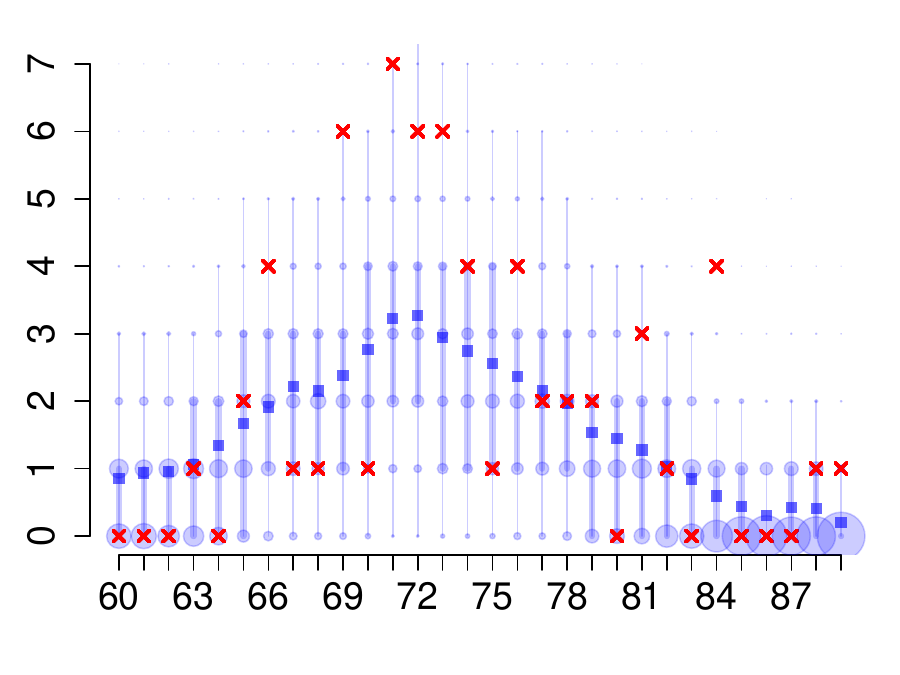} \\
AD-FE & AD-AR & TD-AR \\
\includegraphics[width=0.3\textwidth]{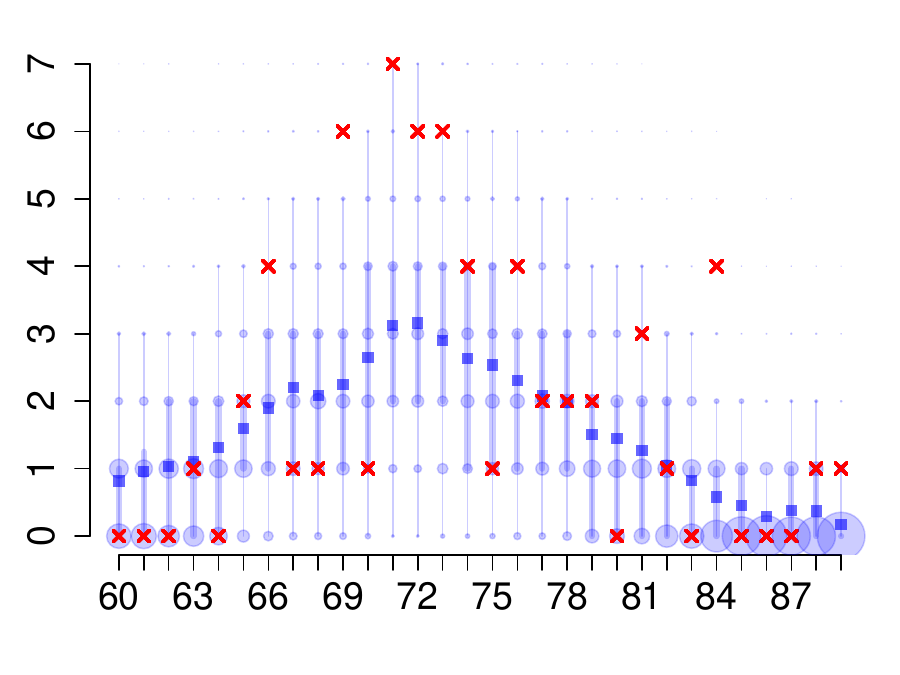} &
\includegraphics[width=0.3\textwidth]{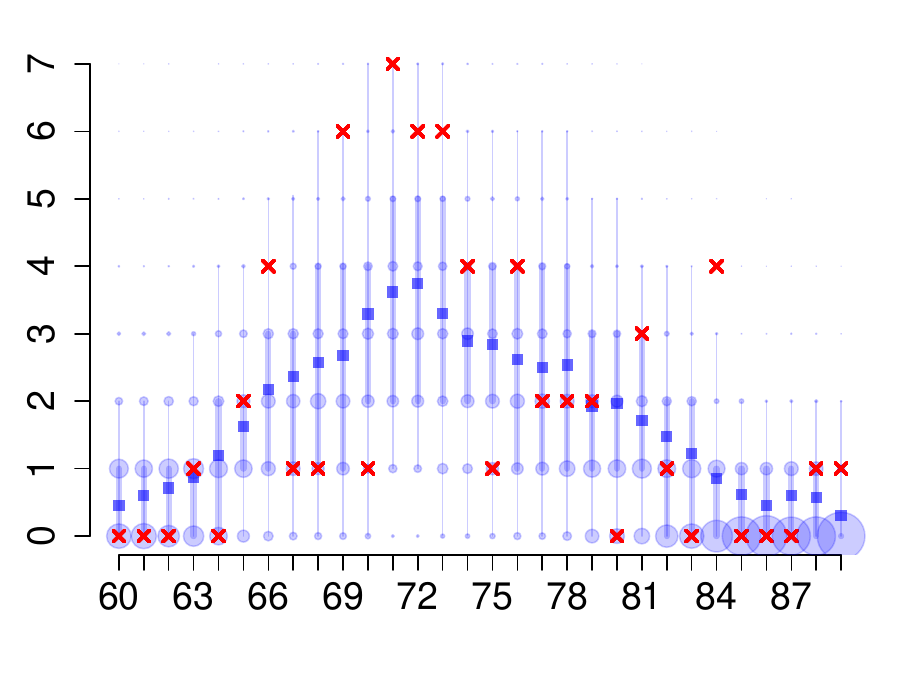} & \\
TD-GP & GP-S1 & \\

\end{tabular}

\end{figure}

\begin{figure}[!ht]
\caption{Posterior distributions for (i) $\mu^{i}$ and $\rho^i$ under model AD-AR (first row); (ii) $\sigma^i$ and $\phi_{ag}^i$ under model AD-GP (second row); (iii) $\rho^i$ under model TD-AR (third row); (iv) $\sigma^i$ and $\phi_{yr}^i$ under model TD-GP (bottom row). All results are for Males and BRA reference population using data from pension fund 1. \label{fig:hiperpost_appendix}}
\centering
\medskip
\begin{tabular}{lcc}
AD-AR: & \raisebox{-.5\height}{\includegraphics[width=0.3\textwidth]{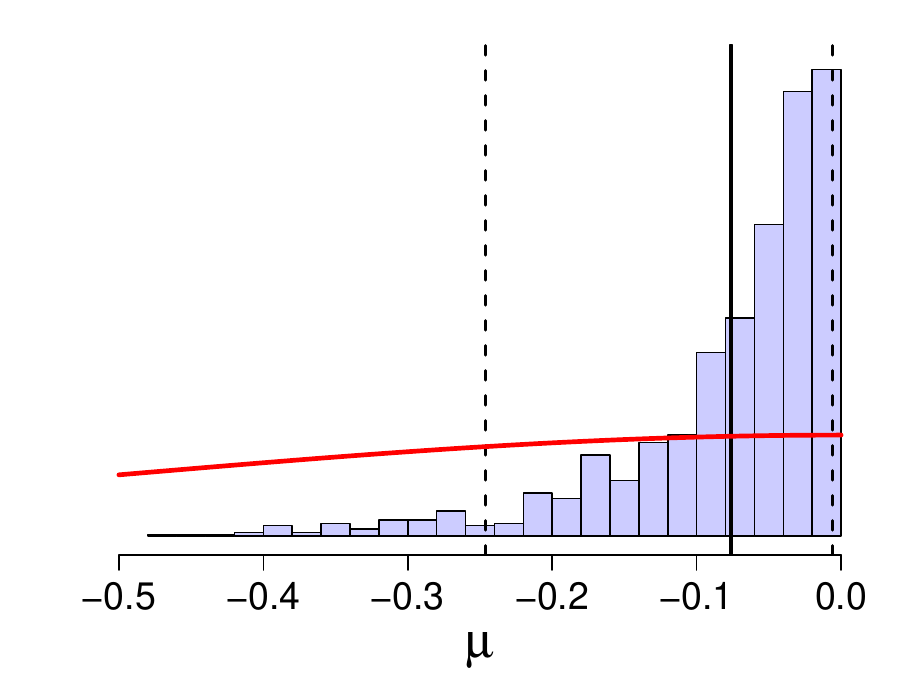}} &
\raisebox{-.5\height}{\includegraphics[width=0.3\textwidth]{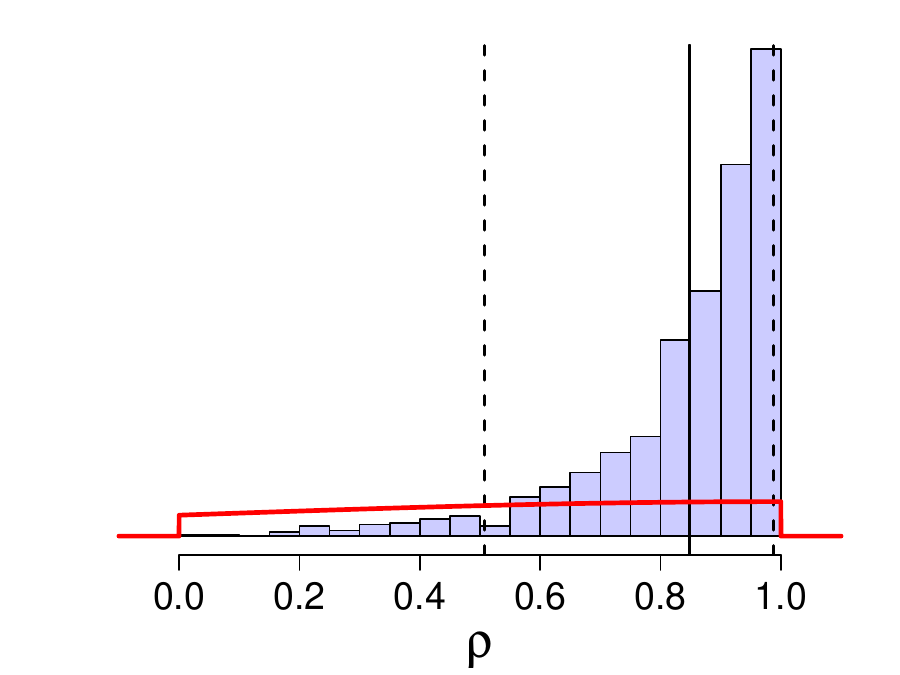}}  \\
AD-GP: & \raisebox{-.5\height}{\includegraphics[width=0.3\textwidth]{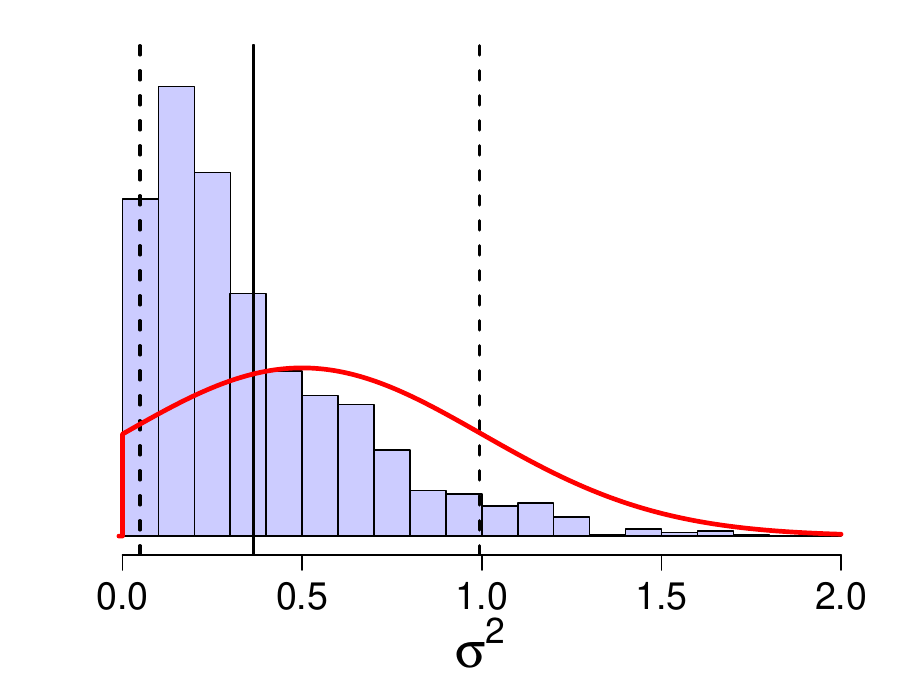}} &
\raisebox{-.5\height}{\includegraphics[width=0.3\textwidth]{length_scale_ag_adgp.pdf}} \\
TD-AR: & \raisebox{-.5\height}{\includegraphics[width=0.3\textwidth]{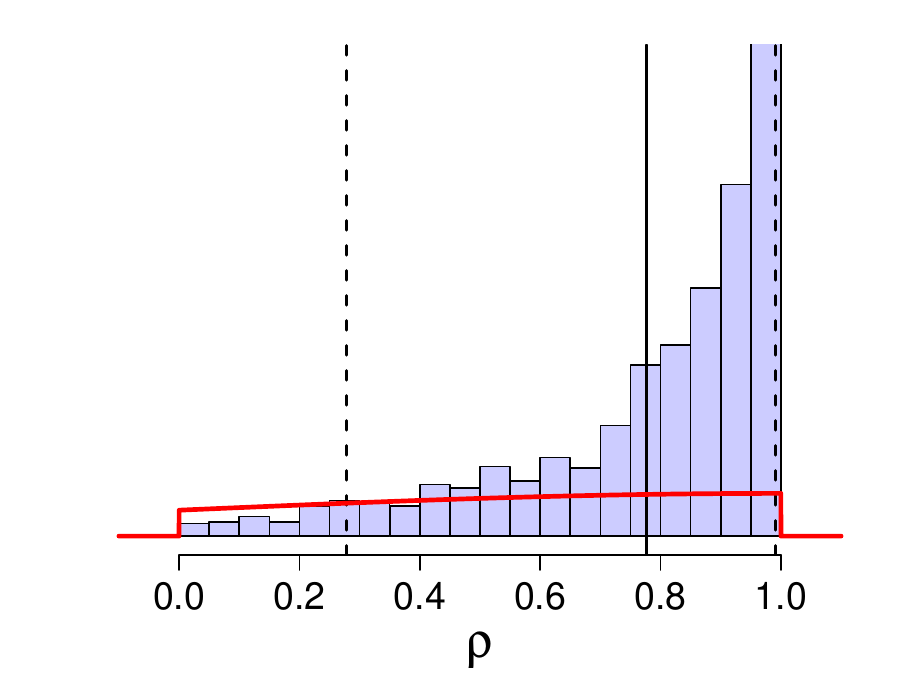} } \\
TD-GP: & \raisebox{-.5\height}{\includegraphics[width=0.3\textwidth]{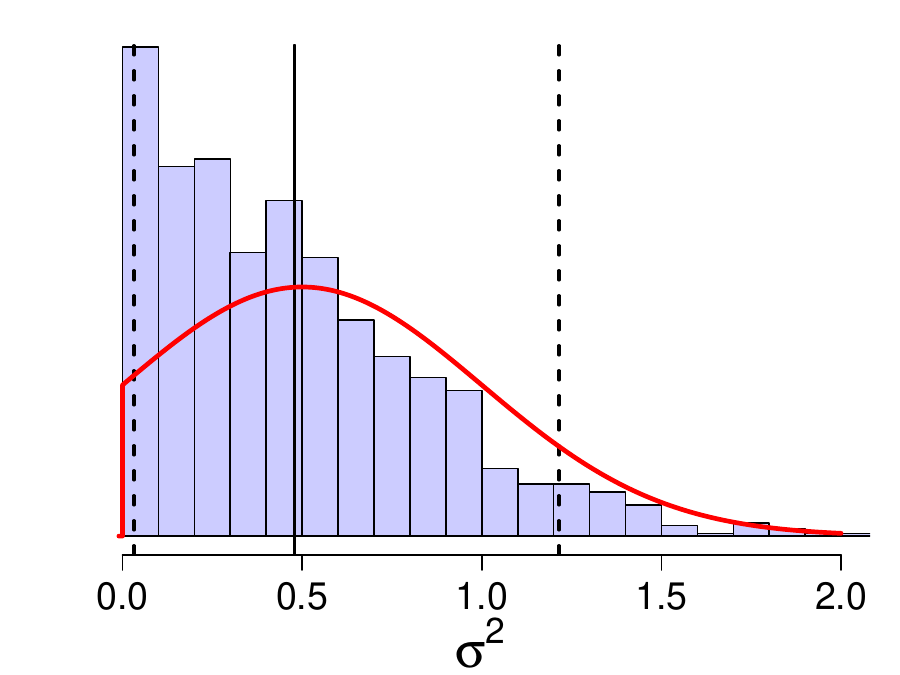}} &
\raisebox{-.5\height}{\includegraphics[width=0.3\textwidth]{length_scale_yr_tdgp.pdf}}
\end{tabular}
\end{figure}

\begin{figure}[!ht]
\caption{Posterior distributions for the hyperparameters $\beta_{0}$ (top row), $\beta_{ag}$ (middle row) and $\phi_{ag}$ (bottom row) under models GP-S1 and GP-S2 for Males. We also include distributions of $\beta_{yr}$ and $\phi_{yr}$ for GP-S2. Vertical lines indicate the posterior means and the dashed lines the posterior 95\% quantile interval. \label{fig:parameters_gps1_gps2}}
\centering
\hspace*{-0.25in}
\begin{tabular}{rlcc}

\includegraphics[width=0.3\textwidth,trim=0.2in 0in 0.2in 0in]{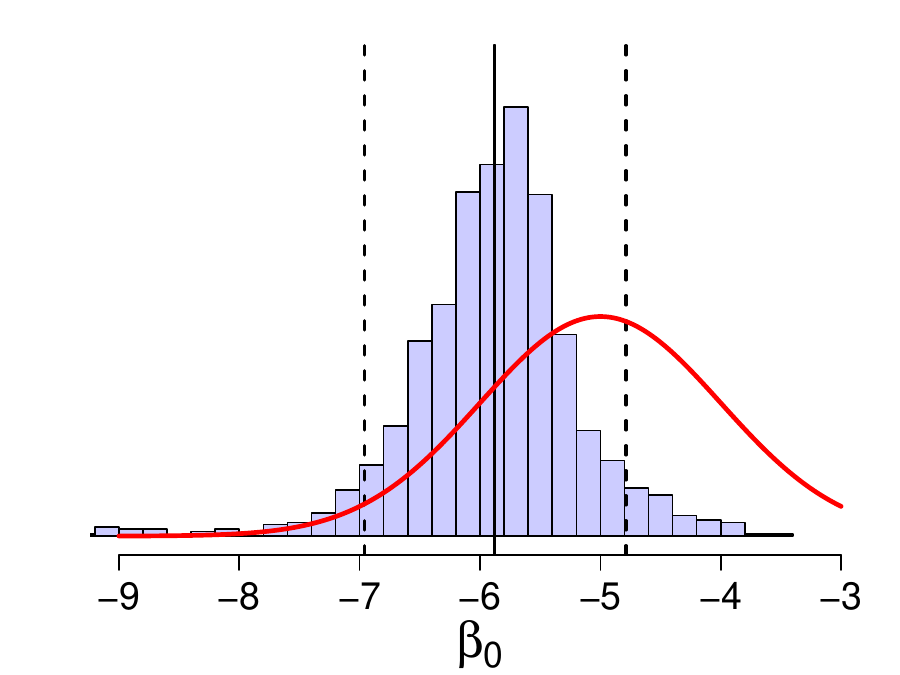} &
\includegraphics[width=0.3\textwidth, trim=0.2in 0in 0.2in 0in]{beta_0_gps1.pdf} \\

\includegraphics[width=0.3\textwidth, trim=0.2in 0in 0.2in 0in]{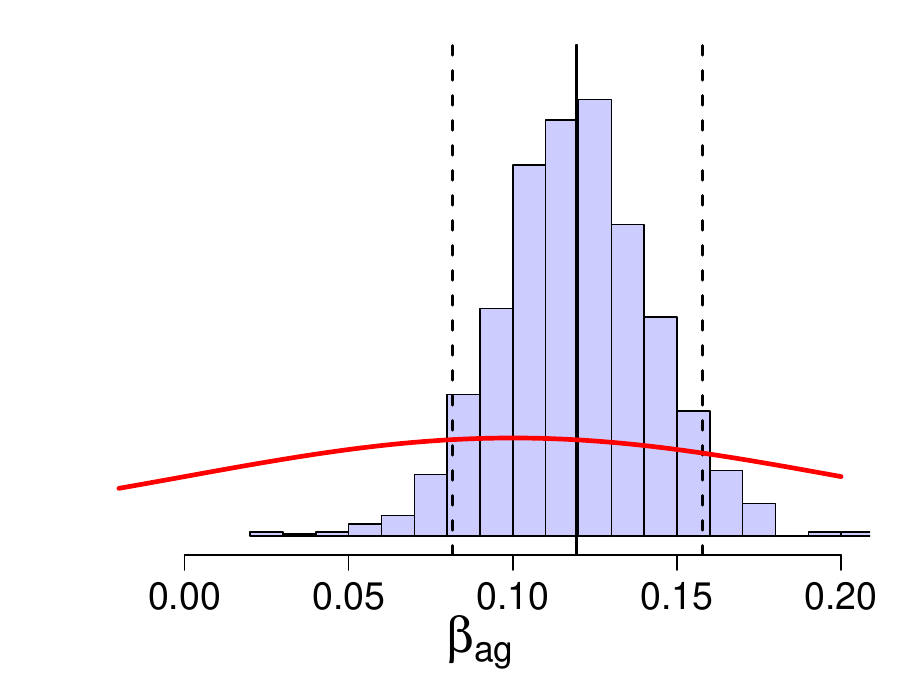} & 
\includegraphics[width=0.3\textwidth, trim=0.2in 0in 0.2in 0in]{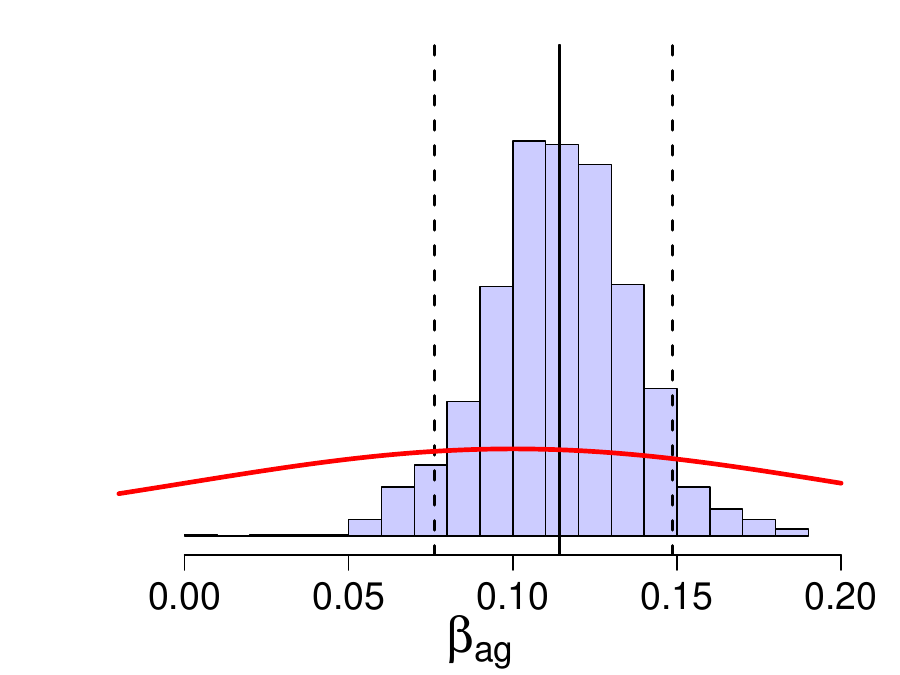}  & $\quad$
& \includegraphics[width=0.3\textwidth,trim=0.2in 0in 0.2in 0in]{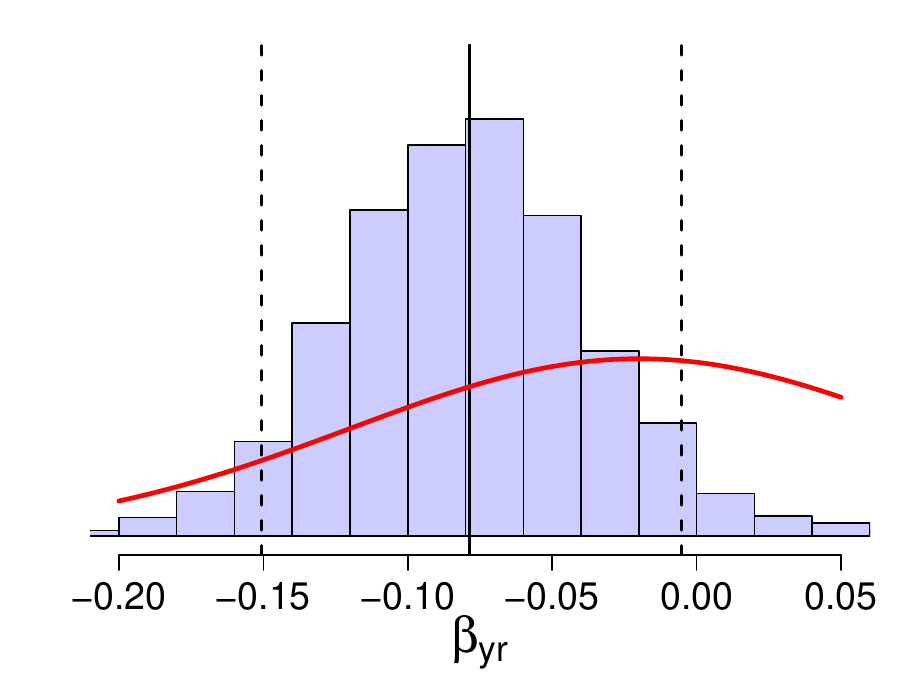}  \\

\includegraphics[width=0.3\textwidth, trim=0.2in 0in 0.2in 0in]{length_scale_ag_gps1.pdf} & 
\includegraphics[width=0.3\textwidth, trim=0.2in 0in 0.2in 0in]{length_scale_ag_gps2.pdf}  
& $\quad$ & \includegraphics[width=0.3\textwidth, trim=0.2in 0in 0.2in 0in]{length_scale_yr_gps2.pdf} \\
GP-S1 & \multicolumn{3}{c}{GP-S2} 
\end{tabular}
\end{figure}

\begin{figure}[!ht]
\caption{Inferred deflators $\theta^i(\cdot)$ for Males and IND reference population across six models. For the FD-1 model we show the prior and posterior densities. 
For all other models, thicker error bars denote the 50\% posterior credible interval, thinner bars the 90\% interval, and the dots the posterior mean. For models AD-FE, AD-AR and AD-GP the horizontal axis denotes age, while for TD-AR and TD-GP it denotes year. For the time-dependent models, 2019 is a forecast.}
\label{fig:posterior_theta_ind}
\centering
\medskip

\begin{tabular}{ccc}
\includegraphics[width=0.3\textwidth]{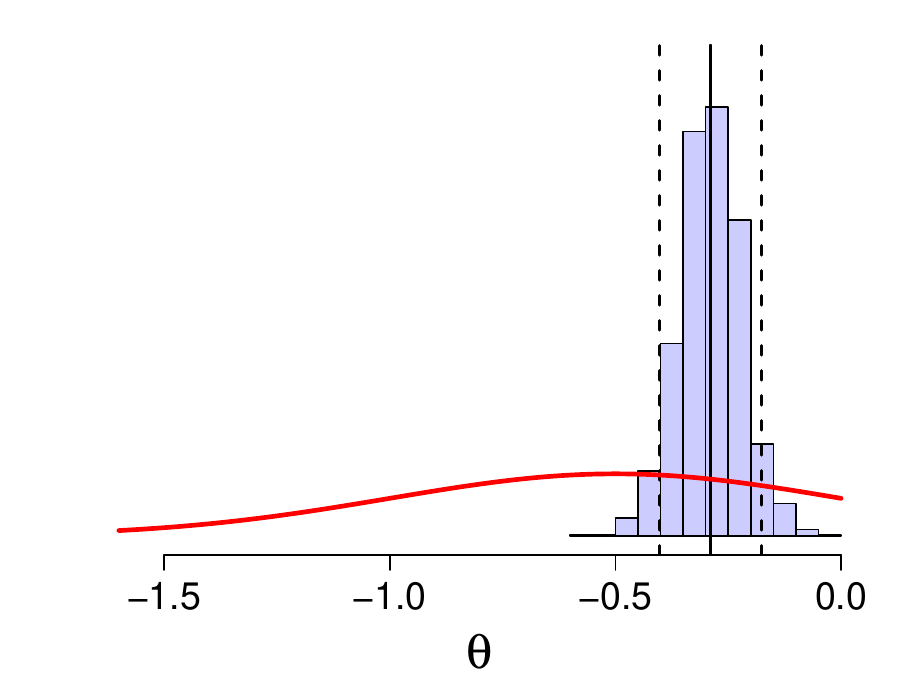}  &
\includegraphics[width=0.3\textwidth]{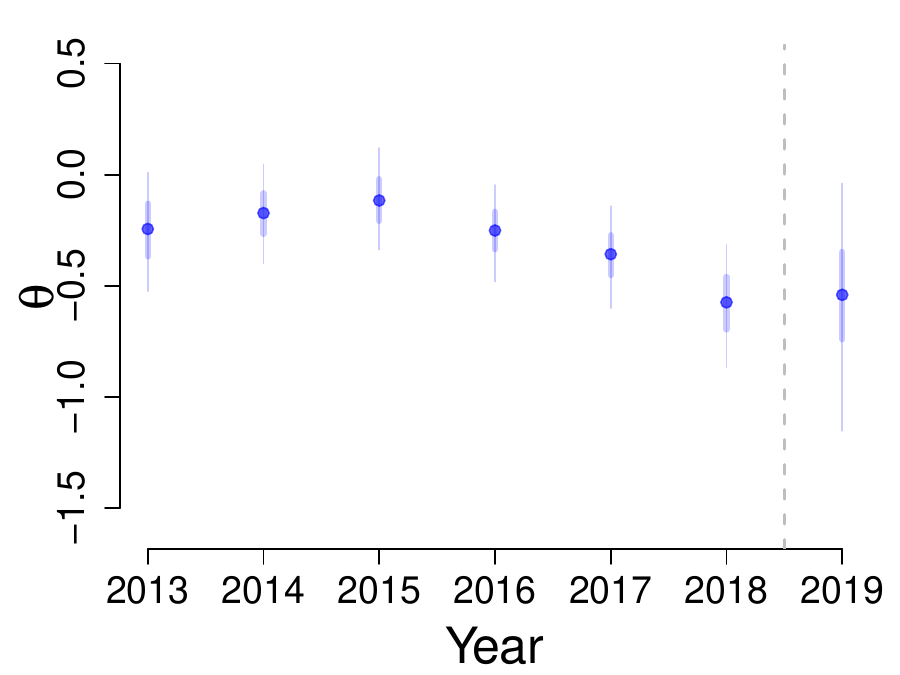} &
\includegraphics[width=0.3\textwidth]{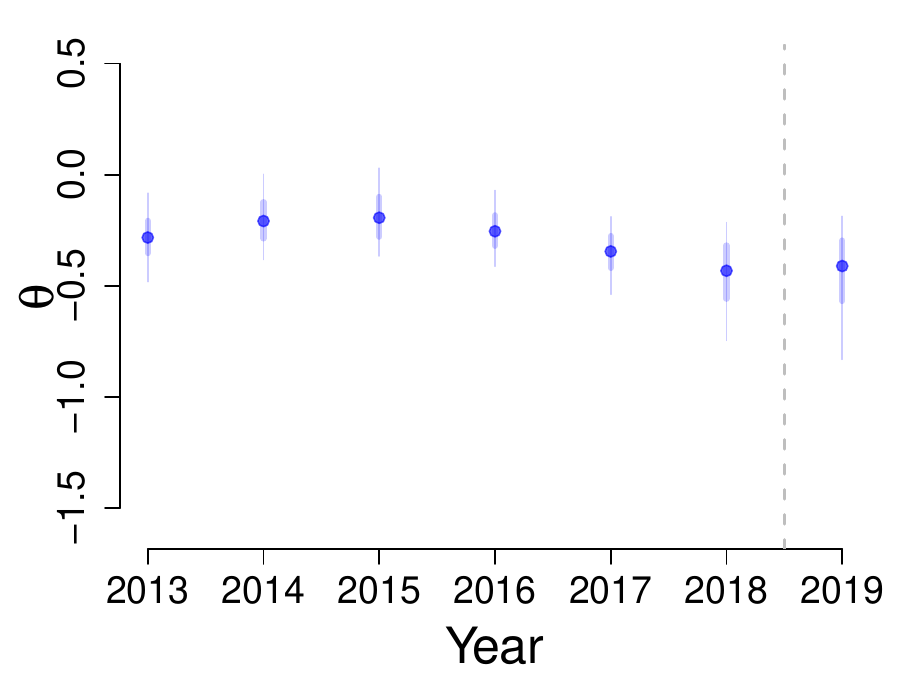} \\
FD-1 & TD-AR & TD-GP \\
\includegraphics[width=0.3\textwidth]{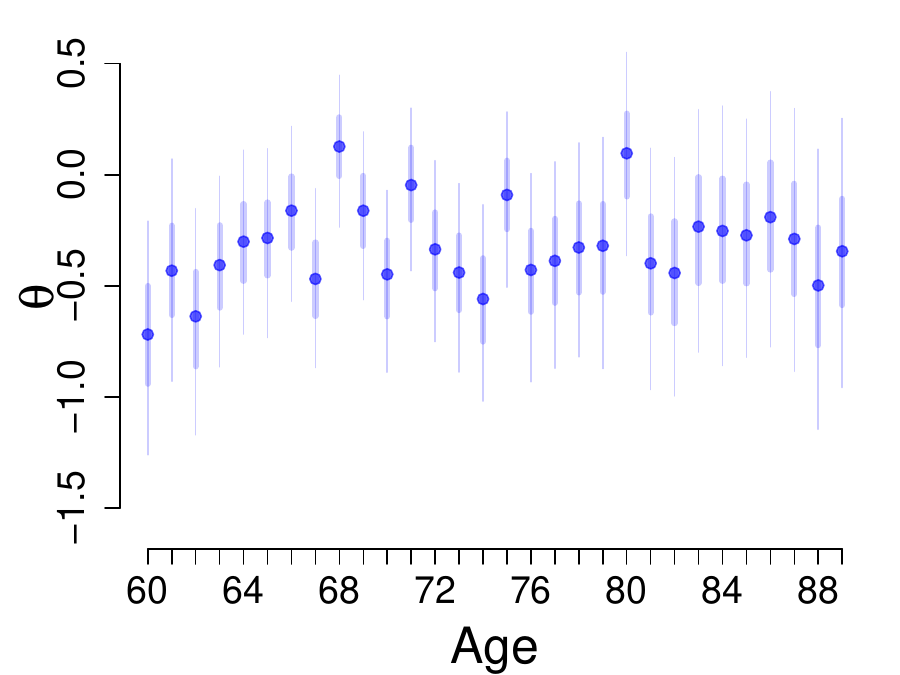} &
\includegraphics[width=0.3\textwidth]{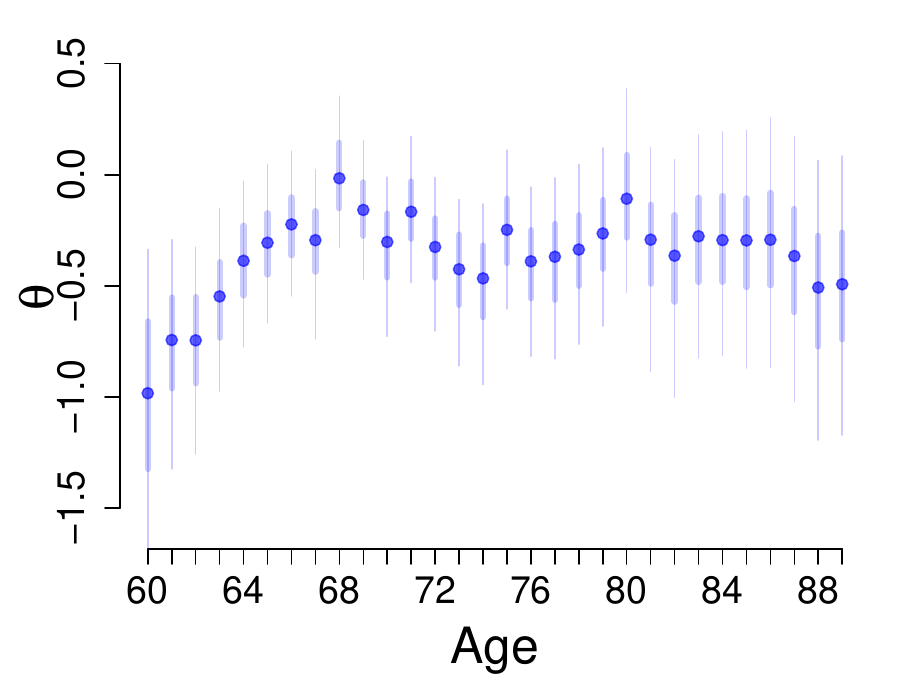} & 
\includegraphics[width=0.3\textwidth]{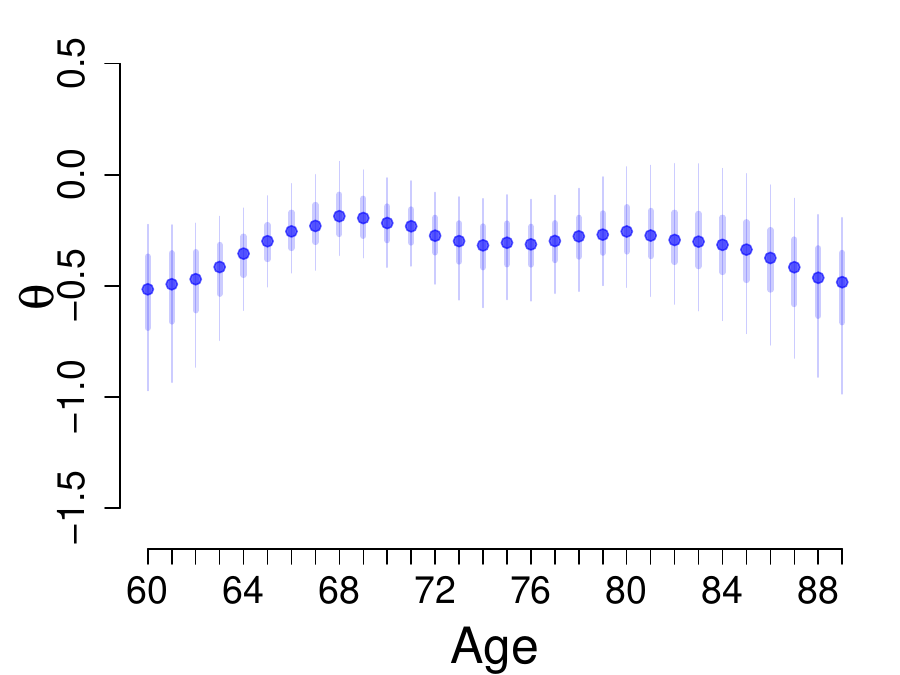} \\
AD-FE & AD-AR & AD-GP \\
\end{tabular}

\end{figure}

\clearpage

\section{Analysis for pension fund 2} \label{sec:alternative_pension_fund}

The alternative pension fund 2 is about half as big as in the first case study. In 2019 for ages 60-89 it included $1,499$ male retirees, with $39$ deaths recorded. Compared to pension fund 1, this subpopulation has a higher mortality:  using $BRA$ as reference population, the mean deflator (based on fitting the FD-1 model) is $\hat{\theta} = -0.29$ (whereas for pension fund 1 $\hat{\theta} = -0.50$, cf.\ Section \ref{sec:deflators}). 

\begin{figure}[!ht]
\caption{Inferred deflators $\theta^i(\cdot)$ for Males in pension fund 2. We use BRA reference population and fit six models. For the FD-1 model we show the prior and posterior densities. For all other models, thicker error bars denote the 50\% posterior credible interval, thinner bars the 90\% interval, and the dots the posterior mean. For models AD-FE, AD-AR and AD-GP the horizontal axis denotes age, while for TD-AR and TD-GP it denotes calendar year. For the time-dependent models TD-AR and TD-GP, 2019 is a forecast.}
\label{fig:posterior_theta_alternative}
\centering
\medskip

\begin{tabular}{ccc}
\includegraphics[width=0.3\textwidth]{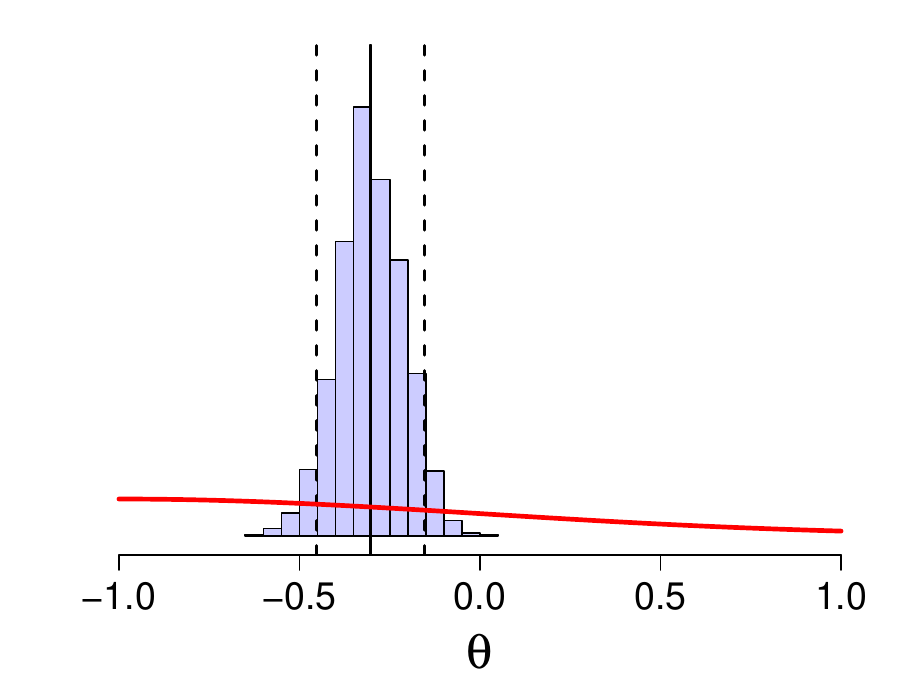}  &
\includegraphics[width=0.3\textwidth]{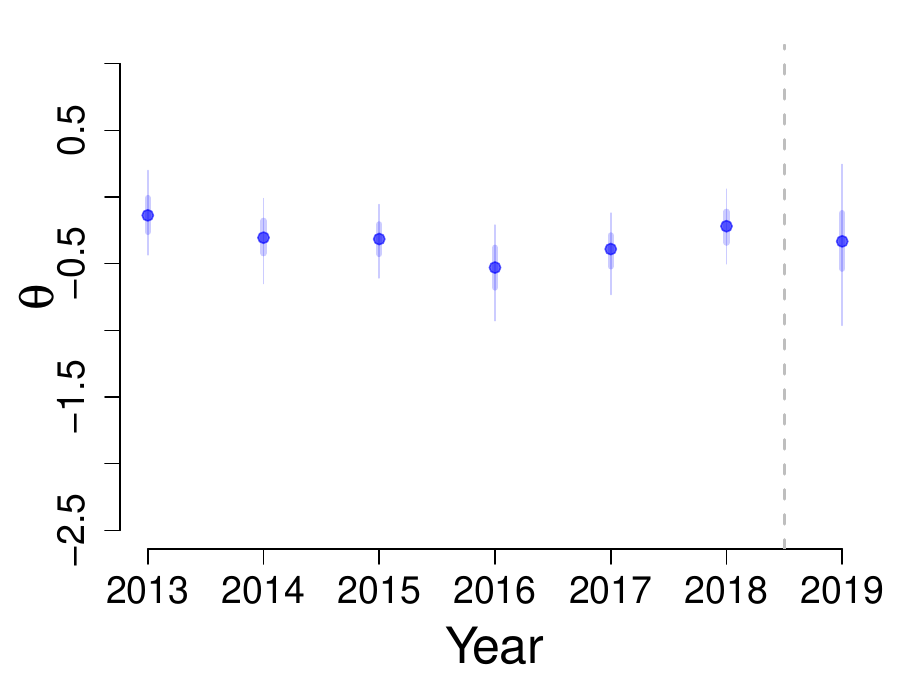} &
\includegraphics[width=0.3\textwidth]{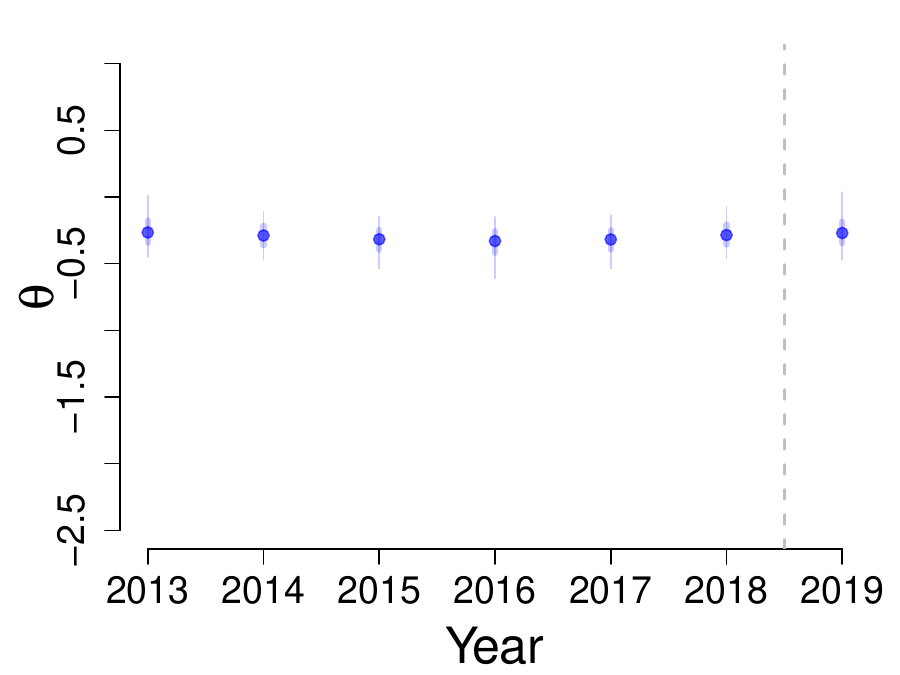} \\
FD-1 & TD-AR & TD-GP \\
\includegraphics[width=0.3\textwidth]{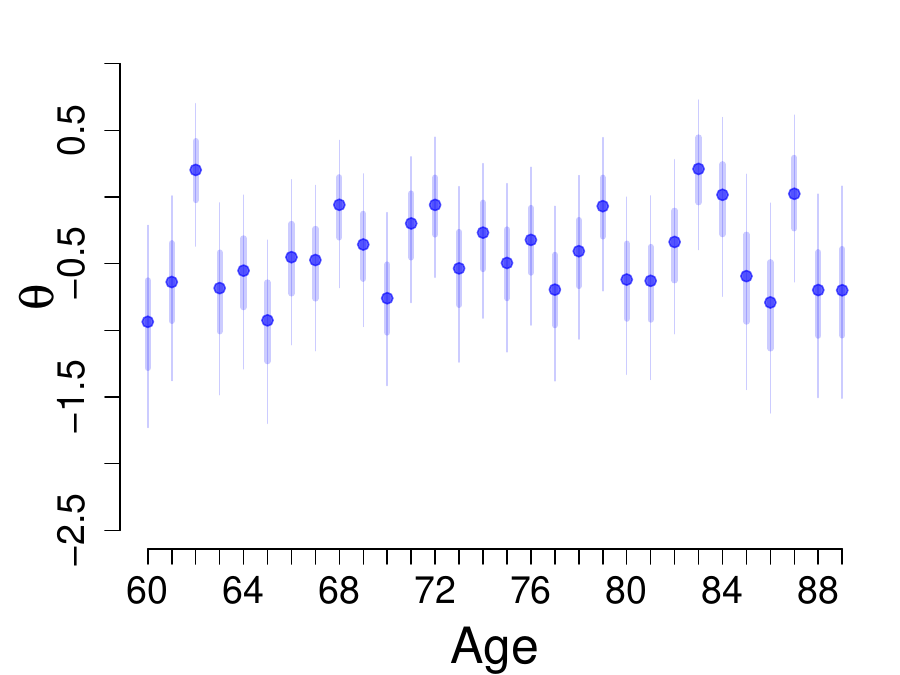} &
\includegraphics[width=0.3\textwidth]{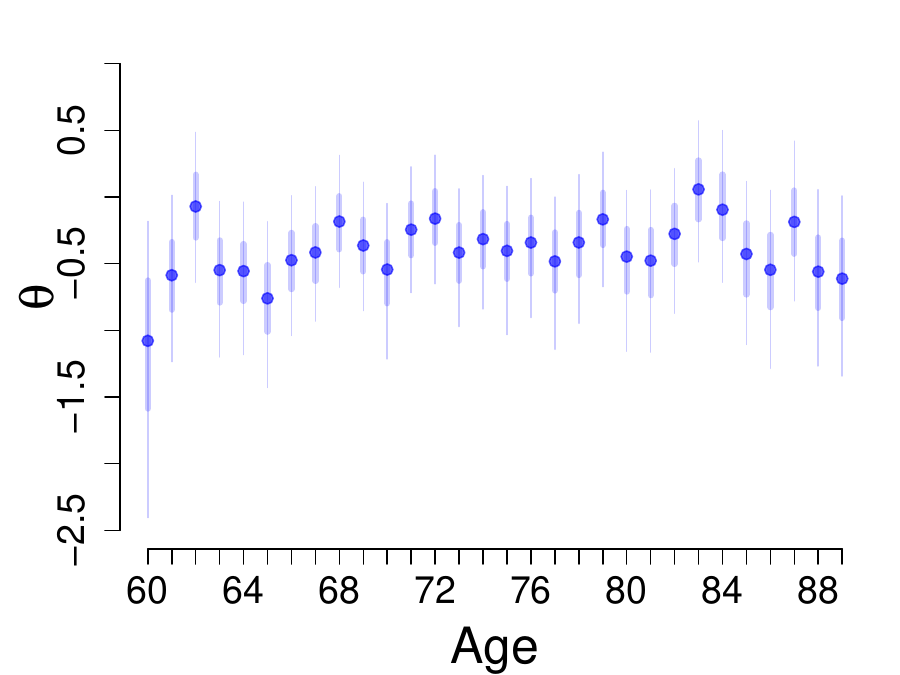} & 
\includegraphics[width=0.3\textwidth]{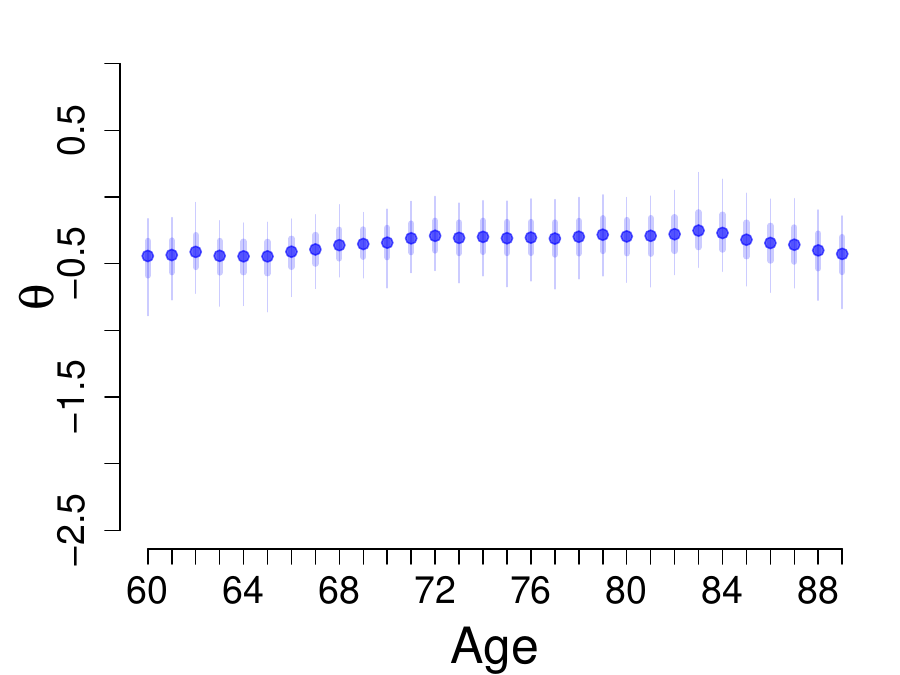} \\
AD-FE & AD-AR & AD-GP \\
\end{tabular}

\end{figure}

In analogue to Figure \ref{fig:posterior_theta}, Figure \ref{fig:posterior_theta_alternative} presents the inferred deflators for the models fitted to this subpopulation. Apart from generally higher deflators compared to the fund analyzed in the main body, one striking difference is much flatter age-structure, with the deflators being almost constant (see the AD-GP results) across the entire age range. We also note that due to the smaller population size, the non-GP AD-FE and AD-GP models are extremely noisy, and their inferred age-specific deflators ``bounce'' around in an undesired manner. This is a telltale sign of overfitting and reaffirms the importance of smoothing across ages that is offered by the GP models. Moreover, the uncertainty around the AD-FE and AD-AR estimates is very large (error bars lengths exceed 1 in Figure \ref{fig:posterior_theta_alternative}) so these models are unable to learn much due to the small population size; in contrast the AD-GP uncertainty is significantly lower and on par with the uncertainty of FD-1.  These features are confirmed in Table \ref{tab:modelsBinNeg_alternative}, where AD-FE and AD-AR perform worse (out-of-sample) than even the constant-deflator FD-1 model, and AD-FE strongly overfits (large gap between in-sample and out-of-sample scores). 

\begin{figure}[ht!]
\caption{Predicted 2019 log-mortality rates for pension fund 2 for Ages 60-89 based on the proposed GP-based models trained on 2013-2018.
AD-GP and TD-GP use $BRA$ as the reference population. Crosses indicate the observed raw mortality rates of pension fund 2 in 2019.}
\centering
\includegraphics[width=0.8\textwidth]{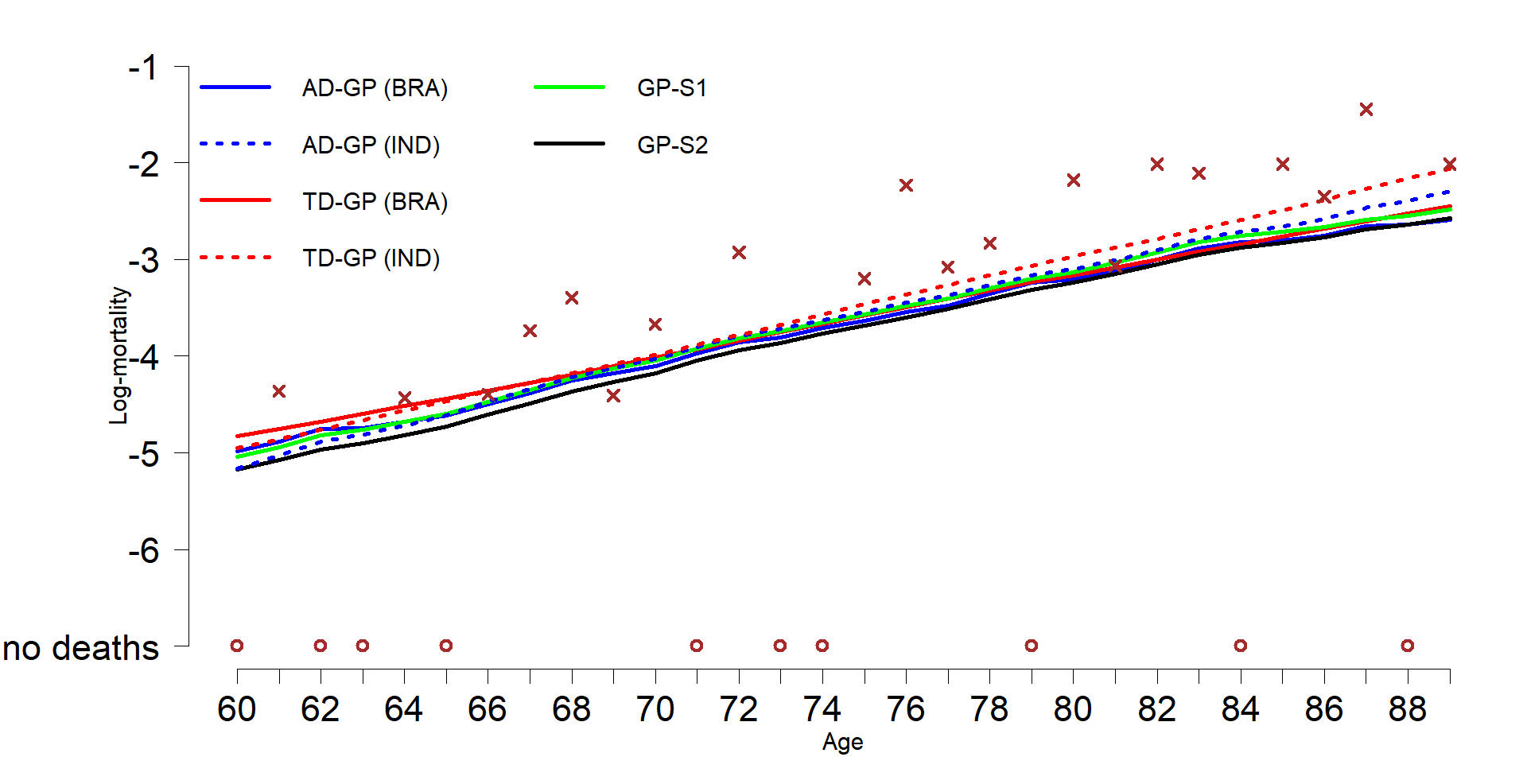}
\label{fig:Fitted_mort_curves_alternative}
\end{figure}

Figure \ref{fig:Fitted_mort_curves_alternative} presents the fitted age-structure of log-mortality of pension fund 2 across the AD-GP, TD-GP and GP-S1/S2 models, compare to  Figure \ref{fig:Fitted_mort_curves}. Although at a first glance the inferred curves appear to underestimate the observed log-mortality, note the many ages where no deaths were observed at all, which ``pull'' down the fitted curves.

Pension fund 1, discussed in the main text, experienced substantial longevity gains in the analysis period (of more than 7\% annually). Since such gains are not sustainable in the long run, our prior choices for the parameters governing mortality improvement were made  to ``tame'' these observed gains. As a consequence, though, both the in- and out-of-sample performance of time-dependent models was negatively impacted (see Table~\ref{tab:modelsBinNeg}). The  population of pension fund 2 does not present similar longevity gains (we estimate a 1-2\% annual gain, e.g., $\beta_{yr} = 0.016$), which provides an opportunity for time dependent models to deliver improved performance. Indeed, in  Table~\ref{tab:modelsBinNeg_alternative}, the TD-GP model achieves the best out-of-sample RPS and log-score metrics. 

As different subpopulations feature different mortality patterns, it should be understood that no single model will ever be the best for all datasets and different specifications should always be tested and compared. Nevertheless, the aggregate results across the two pension funds showcase that GP-based models outperform other deflator approaches: AD-GP being the recommended choice for pension fund 1 and TD-GP for pension fund 2. We also see that the deflator-based methods tend to do better than the direct approach, as GP-S2 continues to overfit and GP-S1 offering a generally middling performance.

\begin{table}[!ht]
\centering
\begin{tabular}{ c|c|c|c|c  }
\hline   

\multirow{2}{*}{Model} & \multicolumn{2}{c|}{RPS}  & \multicolumn{2}{c}{Log-score} \\ \cline{2-5}
 & Out-of-sample & In-sample & Out-of-sample & In-sample \\ \hline

FD-1   & 0.5171 & 0.5386 & 1.2560 & 1.2894 \\

AD-FE  & 0.5279 & 0.4991 & 1.2792 & {\bf 1.2205} \\

AD-AR  & 0.5250 & 0.5060 & 1.2673 & 1.2346 \\

AD-GP  & 0.5197 & 0.5285 & 1.2642 & 1.2691 \\

TD-AR  & 0.5169 & 0.5357 & 1.2593 & 1.2831 \\

TD-GP  & {\bf 0.5147} & 0.5364 & {\bf 1.2557} & 1.2878 \\

GP-S1  & 0.5185 & 0.5291 & 1.2608 & 1.2766 \\
GP-S2  & 0.5199 & {\bf 0.4952} & 1.2638 & 1.2252 \\ \hline 

\hline
\end{tabular}
\caption{Mean of the yearly performance indexes \eqref{eq:log-score}-\eqref{eq:rps} for leave-one-out cross-validation across years 2013-2019 for pension fund 2. The shown metrics are for ages 60-89 for Males and $BRA$ as reference population, except for models GP-S1 and GP-S2. Out-of-sample and in-sample results considering models described in Table \ref{tab:modSOGP} and Negative Binomial likelihood. Bolded numbers indicate the best-performing model for each metric. \label{tab:modelsBinNeg_alternative}}
\end{table}

\end{document}